\documentclass{aa}  

\usepackage{graphicx}
\usepackage{subfig}
\usepackage{txfonts}
\usepackage{xcolor}
\usepackage{tabularx}
\usepackage{csvsimple}
\definecolor{seagreen}{rgb}{0.190, 0.525, 0.361}
\definecolor{darksalmon}{rgb}{0.914, 0.588, 0.478}
\definecolor{steelblue}{rgb}{0.274 0.510 0.706}
\definecolor{purple}{rgb}{0.5 0.0 0.5}
\definecolor{darkred}{rgb}{0.8 0.0 0.0}
\definecolor{darkgold}{rgb}{0.8 0.543 0.0}
\definecolor{disagreeablegray}{rgb}{0.00 0.00 0.00}

\newcommand{\REFEREE}[1]{\textcolor{disagreeablegray!100}{#1}}

\begin{document} 

   \title{Interpreting automatic AGN classifiers with saliency maps}

   \author{T. Peruzzi \inst{1}\thanks{tobia.peruzzi@studenti.unipd.it}
          \and
          M. Pasquato\inst{2,3,4}            \thanks{mp5757@nyu.edu}
          \and
          S. Ciroi\inst{1,3}
          \and
          M. Berton\inst{5,6}
          \and
          P. Marziani\inst{3}
          \and
          E. Nardini\inst{7,8}
          }

   \institute{Dipartimento di Fisica e Astronomia ‘G. Galilei’, Università degli Studi di Padova, Vicolo dell’Osservatorio 3, 35122, Padova (Italy)
            \and
            Center for Astro, Particle and Planetary Physics (CAP$^3$), New York University Abu Dhabi 
            \and
             INAF, Osservatorio Astronomico di Padova, Vicolo dell'Osservatorio 5, I--35122 Padova, Italy
            \and  
             INFN- Sezione di Padova, Via Marzolo 8, I--35131 Padova, Italy
            \and
             Finnish Centre for Astronomy with ESO (FINCA), University of Turku, Vesilinnantie 5, 20014, Turku, Finland;
            \and
             Aalto University Metsähovi Radio Observatory, Metsähovintie 114, 02540, Kylmälä, Finland
            \and
             Dipartimento di Fisica e Astronomia, Universit\`a di Firenze, via G.Sansone 1, I-50019 Sesto Fiorentino, Firenze, Italy
            \and
             INAF, Osservatorio Astrofisico di Arcetri, Largo Enrico Fermi 5, I-50125 Firenze, Italy.
             }

   \date{Received    ; accepted    }

 
  \abstract{Currently, classification of the optical spectra of active galactic nuclei (AGN) into different types is based on features such as line widths, intensity ratios, etc. While well founded on AGN physics, this approach involves some degree of human oversight and cannot scale to large data sets. Machine learning (ML) tackles such a classification problem in a fast and reproducible way, but is often -and not without reason- perceived as a black box. However, ML interpretability and explainability are active research areas in computer science, increasingly providing us with tools to alleviate this issue. We apply ML interpretability tools to a classifier trained to predict AGN type from spectra. Our goal is to demonstrate the use of such tools in this context, obtaining for the first time insight into an otherwise black box AGN classifier. In particular we want to understand which parts of each spectrum most affect the predictions of our classifier, checking that the results make sense in the light of our theoretical expectations. We train a support-vector machine on $3346$ high-quality, low redshift AGN spectra from SDSS DR15. We consider either two-class classification (type 1 VS 2) or multiclass (type 1 VS 2 VS intermediate). The data set was previously and independently hand-labeled into types 1, 2 and intermediate i.e. sources in which the Balmer line profile consists of a sharp, narrow component superimposed on a broad component. We perform a train-validation-test split, tuning hyperparameters and independently measuring performance with a variety of metrics. On a selection of test-set spectra, we compute the gradient of the predicted class probability at a given spectrum. Regions of the spectrum are then color-coded based on the direction and the amount by which they influence the predicted class, effectively building a saliency map. We also visualize the high-dimensional space of AGN spectra using t-distributed stochastic neighbor embedding (t-SNE), showing where the spectra for which we computed a saliency map are located. Our best classifier reaches an F-score of $0.942$ on our test set (with $0.948$ precision and $0.936$ recall). We compute saliency maps on all misclassified spectra {in the test set} and on a sample of randomly selected spectra. Regions that affect the predicted AGN type often coincide with physically relevant features, such as spectral lines. t-SNE visualization shows good separability of type 1 and type 2 spectra. Intermediate-type spectra either lie in-between as expected or appear mixed with type 2 spectra. Misclassified spectra are typically found among the latter. Some clustering structure is apparent among type 2 and intermediate-type spectra, though this may be an artifact. Saliency maps show why a given AGN type was predicted by our classifier resulting in a physical interpretation in terms of regions of the spectrum that affected its decision, making it no longer a black box. These regions coincide with those used by human experts such as relevant spectral lines, and are even used in a similar way, with the classifier e.g. effectively measuring the width of a line by weighing its center and its tails oppositely.}


   \maketitle

\section{Introduction}
Active galactic nuclei (AGN) are the most energetic non-transient phenomena in the Universe. AGN can be found in nuclei of galaxies characterized by highly ionized gas not correlated with stellar activity. The gas surrounding the AGN can be photoionized by photons produced by accretion mechanisms onto a supermassive black hole (SMBH), with $M_{BH} \approx 10^{6} - 10^{9} M_{\odot}$, which accretes material from the surrounding interstellar medium \citep[][]{salpeter64,zeldovichnovikov65,1969Natur.223..690L, 1984ARA&A..22..471R}.

Classically AGNs, and in particular Seyfert galaxies, are divided into two groups { \citep[][]{1971Ap......7..231K, khachikianweedman74}: Type 1 and Type 2.
The corresponding physical interpretation -the so-called Unified Model- is that for Type 1 the observer looks directly into the unobscured accretion disk surrounded by fast-moving gas clouds, and for Type 2 the line-of-sight into the accretion disk is blocked by an obscuring medium \citep[][]{1993ARA&A..31..473A, 1995PASP..107..803U}}.

{AGNs emit radiation in virtually all bands and consequently have historically been described in terms of several classes of objects depending on the band they were discovered in. AGN classification is reviewed in detail by \cite{2017A&ARv..25....2P}, which also includes a systematic discussion of the definitions of different but sometimes overlapping classes of AGNs defined over time by observational astronomers in bands ranging from the radio to the gamma-rays, the so-called AGN zoo. In the following we focus on the problem of classification into Type 1 and Type 2 because of its implications for statistical analyses \citep[see][which also discusses a refinement of the original unified model]{2012ApJ...747L..33E} on large catalogs and to illustrate an application of ML interpretability tools.}

Classification of sources into Type 1 and Type 2 is typically based on features observed in the optical spectrum, such as the full width at half maximum (FWHM) of the broad H$\beta$ line: type 1 AGN are classically defined as having FWHM of  permitted "broad"  lines in excess to the ones of forbidden lines that rarely exceed 1000 km/s, generally accompanied by an intense, blue continuum; type 2 show permitted and forbidden emission lines with comparable width \citep{khachikianweedman74}. Among type 1 AGN, the ratio of equivalent widths between FeII optical emission and the HI Balmer emission line H$\beta$ helps to classify large samples of  quasars along a main sequence \citep[][]{1992ApJS...80..109B}. This approach may be supplemented by observables in different bands, leading to the so-called four-dimensional Eigenvector 1 (4DE1) parameter space \citep[][]{2000ARA&A..38..521S, 2000ApJ...536L...5S, 2007ApJ...666..757S}.
These methods of AGN classification are firmly grounded in our understanding of AGN physics but are hard to automate, requiring at least some human oversight. {Direct quantification of the classification performance attained by humans is obviously hard, as it would involve setting up a controlled classification experiment, but there are documented instances of spurious source identifications which were overturned on closer inspection, e.g. \citep[][]{2020A&A...636L..12J}. The performance of automatic approaches on the other hand can be easily evaluated on an unseen test set.  For these reasons}, AGN classification for extremely large data sets, such as the Sloan Digital Sky Survey (SDSS), is likely to require an automated approach. The challenge we are facing is to make classification fast and accurate, without turning the classification process into a black box and losing physical interpretability. 

{Surprisingly, automatic ML classification of AGN optical spectra was attempted relatively few times based on artificial neural networks \citep[][]{1996PASA...13..207R, 2014A&A...567A..92G} and nearest neighbor schemes \citep[][]{2007ChA&A..31..352Z} only. In all cases the focus was only on correct automatic classification rather than on the interpretability of the resulting model. This is also the case for the most recent and to our knowledge most accurate AGN classification result based on a supervised ML framework presented by \cite{2020ASPC..522..421T}. They trained various black-box machine learning models on $10000$ SDSS DR-14 spectra, achieving remarkably high classification performance ($\approx 93\%$ in terms of the F-score metric, which we will discuss later). 
The authors also use random forest feature importance to gain some insight into which principal components of the feature space of spectra are more informative, but do not further discuss their physical meaning.
Notwithstanding their great classification performance, the current state of the art in automated AGN classification lacks interpretability: how are these models achieving such high performance?}
{In the following we will focus on this question, while pointing the reader interested in a general discussion of ML in astronomy to the excellent review by \cite{2020WDMKD..10.1349F}.}


{Interpretability and explainability are open research areas in machine learning, and a variety of techniques have been proposed depending on the context in which the need for model explanation arises \citep[see][for a review]{molnar2019interpretable}. In astronomy and science in general, the ability to provide an explanation in addition to a bare  prediction is likely crucial for adoption of ML methods.}

{While interpretability techniques are increasingly being applied to a variety of astronomical problems \citep[see e.g.][]{2019ApJ...882L..12P, 2020arXiv200614305V, 2020arXiv200511066Z}, alongside with natively interpretable models such as simple decision trees \citep[e.g.][]{2019MNRAS.485.5345A}, they are still far from the norm in the field.}
{Generally speaking, interpretability tools are either model-specific or model agnostic. The former apply only to a specific set of ML models, while the latter potentially apply to any model, including a black-box one; these are clearly more interesting for application to astronomy. In the following we will visualize the gradient of our classifier's prediction (more precisely the relative change in the predicted probability or confidence for the predicted class), which is applicable to any underlying ML model as long as it is differentiable. The gradient is cheap to compute, clearly indicates how to modify a given instance (an AGN spectrum in our case) to change the associated prediction, and can be readily visualized.}

{In this paper we obtain comparable accuracy to \cite{2020ASPC..522..421T}, also using a support-vector machine \citep[SVM; ][]{cortes1995support}. We then explain our trained classifier's decision on an individual basis by visualizing its gradient by a so-called saliency map \citep[][]{simonyan2013deep} given any AGN spectrum. SVMs are differentiable}, allowing us to compute the gradient of the predicted class probability at any given point in feature space. Since the coordinates of this space are the fluxes measured for each wavelength in our spectra, we can use the gradient computed at any given spectrum to visualize which parts of the spectrum are responsible for a Type 1 classification (slightly increasing the flux at those wavelengths increases the predicted probability of being Type 1), which parts are irrelevant (increasing the flux has no effect), and which parts pull in the opposite direction, towards a Type 2 classification (increasing the flux decreases the predicted probability of being Type 1). This can be conveniently shown as a color-coding of the spectrum under consideration and is an easy way to check what the model is basing its predictions on.

{In addition to interpretability tools applied to classifiers, visualization and visual clustering based on dimensionality reduction approaches where high-dimensional data is mapped to a low-dimensional space such as a plane for visualization purposes are also becoming more commonplace in astronomy \citep[see e.g.][]{2018MNRAS.473.4612K, 2018A&A...619A.125A, 2019arXiv191003085L, 2019amos.confE..17F, 2020ApJ...891..136S, 2020AAS...23544004S, 2020AAS...23517030K}, with applications also to AGNs ranging from time-tested linear methods such as principal component analysis \citep[][]{2004AJ....128.2603Y, 2004AJ....128..585Y} to advanced deep learning approaches \citep[][]{2019ApJS..240...34M, 2020arXiv200210464P}.
Since we are using saliency maps as an instance-by-instance explanation of our ML model it is natural to leverage dimensionality reduction to represent AGN spectra on a plane, where we then show which instances (data points) we are examining. This also allows us to visualize the position of misclassified instances with respect to the other data in our set.}

In Sect. 2 we will describe the dataset used, in Sect 3. the supervised classification setup. In Sect. 4 we will present the SVM performance (Subsec. 4.1), the AGN spectra space visualized with the dimensionality reduction algorithm t-SNE (Subsec. 4.2) and the application of saliency map interpretability tool to AGN spectra (Subsec. 4.3). In Sect. 5 we provide a summary of the results reached in this work.

\section{Data}

Our dataset is composed by $680$ type 1, $2145$ type 2 and $521$ intermediate type AGN spectra {from the SDSS survey. All of them have been accurately classified by previous works in the literature, and are expected to have a lower rate of misclassification than what is typically achieved with unsupervised sample selection \citep[e.g., sect. 2 of][]{Berton20}. For this reason they are well-suited to test our automatic classification procedure.}

{The selection of type 1 spectra is described in detail by \citet{Marziani13}. Firstly, they selected sources cataloged as quasars in the SDSS DR7 in the redshift range 0.4 - 0.75, and with magnitude brighter than 18.5 in $g$, $r$, or $i$ band, to ensure a good spectral quality. They also included sources with FWHM(H$\beta$) $<$ 1000 km s$^{-1}$ selected by \citet{Zhou06}, which usually are not classified as quasars by SDSS. After a visual inspection to remove bad quality spectra, they included 680 sources in the final sample.}

{The selection of the type 2 and intermediate type spectra was carried out by \citet{Vaona2012}. In SDSS DR7 they selected all the sources showing the [O II]$\lambda$3727, [O III]$\lambda$5007, and [O I]$\lambda$6300 lines, with an additional criterion on the signal-to-noise ratio (S/N)([O I]$\lambda 6300) > 3$. This sample of 119226 sources was subsequently reduced by applying a redshift threshold 0.02 $\leq$ z $\leq$ 0.1. The lower limit was needed to ensure the presence of the [O II] line, while the upper limit to avoid contamination from extranuclear sources within the fibre aperture. An empirical criterion based on line ratios suggested by \citet{Kewley06} was applied, to remove sources without AGN activity (see eq. (1) of \citealp{Vaona2012}). The remaining objects were further analyzed on the basis of the diagnostic diagrams by \citet{Veilleux87} and their H$\alpha$ width, and were finally divided into two samples of 2153 Seyfert 2 and 521 intermediate type AGN. Thanks to these strict selection criteria, their spectra had a typical S/N, defined here as the ratio between the mean flux of the 5100\AA continuum and the standard deviation in the same spectral region, between 10 and 40, directly comparable to that of the type 1 sample.}

Intermediate type AGN show Balmer line profiles  consisting of a sharp, narrow component superimposed on a broad component \citep{osterbrockkoski76,osterbrock81,osterbrock91}. Following the classification proposed by Osterbrock, they are distinguished into 1.2, 1.5, 1.8 in order of decreasing prominence {of} the broad component. For the context of the classification task presented in this work, they will be considered as a single type{, since an additional subdivision would require a level of sophistication not necessary at this stage.}. 

Every spectrum has been shifted to rest-frame using the values of $z$ given by the SDSS and normalized to the flux value {at} 
$5100$ \AA in the rest frame. This value is chosen in order to {normalize on} a flux that belongs to the continuum and not to an emission line or some other component.


In order to perform classification on a fixed number of spectral features, we need to turn each spectrum into an array of normalized fluxes of the same length. Every spectrum in the data set has thus been interpolated over $1000$ points at equally spaced wavelengths obtaining flux values at the same wavelengths for every spectrum. Over the range of wavelength overlap this results in an effective resolution in wavelength strictly higher than the nominal SDSS resolution in the same range, therefore no information is lost in the interpolation. These flux values constitute our features, so our feature space is $1000$-dimensional. We restricted the range of our interpolation to the shared overlap of our spectra, i.e. between the maximum among the minimum wavelengths of all the spectra and the minimum among the maximum wavelengths of all the spectra, so that we could include all spectra in the final sample without having to add padding. Note that this approach somewhat reduces the amount of information available to our classifier with respect to that used during human classification, because e.g. some lines used in the latter may end up outside our adopted range. The values of the resulting wavelength range are reported in Table \ref{tab:interp_range}. 

\begin{table}
\caption[Wavelength interpolation ranges.]{Extremes of the wavelength interpolation ranges for type 1, type 2 and intermediate type AGN spectra. \REFEREE{Columns: AGN types in our dataset (first three columns from the left) and adopted range in the last column. Rows: value of minimum and maximum wavelenghts in \AA.} 
}
\label{tab:interp_range}
\centering
\begin{tabular}{lllll}
\hline \hline
 & Type 1 & Type 2 & Int. & Adopted range \\ 
\hline
Min & 2713.93 $\AA$ & 3727.07 $\AA$ & 3728.91 $\AA$ & 3728.91 $\AA$ \\
Max & 5265.95 $\AA$ & 6955.6 $\AA$ & 8318.88 $\AA$ & 5265.95 $\AA$ \\
\hline
\end{tabular}

\end{table}

\section{Supervised classification setup}
{For classifying AGN spectra we selected a support-vector machine (SVM) classifier \citep{cortes1995support} both for two-class classification between type 1 VS type 2 and multiclass with type 1 VS type 2 VS intermediate.} {SVMs look for a maximum margin hyperplane separator between the classes, possibly after an implicit transformation into an higher dimensional space where data that is not linearly separable may become such. Maximizing margin means that the separation surface is as far as possible from any data point, which is an additional constraint with respect to other methods that just find a separation surface. Intuitively this reduces uncertainty in classification (since points are far away from the separation surface, i.e. they are firmly classified) and results in a boundary between classes that depends only on a handful of training data points near the surface, the eponymous support vectors. It was shown empirically that SVMs have good performance on a variety of structured data, text, and other classification tasks \citep[see e.g.][]{manning2008introduction}. In the following we use SVMs in the scikit-learn \citep[][]{scikit-learn} implementation for python. We make use of soft-margin classification, so the separating hyperplane is allowed to make some classification mistakes if this increases margin, but these mistakes are weighed negatively within the cost function that is optimized to train the SVM. The cost of mistakes is a hyperparameter that we fine-tune in validation together with other hyperparameters such as the kernel used for nonlinear SVM, as described in the following.}

The whole dataset was randomly divided into a training and a test set with a $80\% - 20\%$ split. The training set was further randomly split into training and validation sets, again with a $80\% - 20\%$ split, so the final proportions are training-$64\%$, validation-$16\%$, and testing-$20\%$.  The hyperparameter optimization (see below) took place within a $5$-fold cross-validation loop, while the test set was kept as a holdout set from the beginning, i.e. it was not involved in any cross-validation loop.
The train-validation-test split is adopted in order to have a subset to select the best set of parameters for the classifier (the validation set) and a subset of unseen data in order to test the performance of the {best model on unseen data}. The latter is one of the techniques used in ML to avoid overfitting, that {happens} when a ML model is unable to generalize well {to} new data. The random partitioning was unstratified, that is {performed} without imposing any kind of {fixed} ratio between the number of samples belonging to different classes, given the relatively balanced nature of our data set with respect to the different class frequencies. 
However during all training steps of our SVM, we applied weights inversely proportional to class frequency in an attempt to counter class inbalance, using the \emph{class$\_$weight$=$balanced} option in scikit-learn. In Tab.~\ref{tab:train_val_test_sets} we show the frequency of the classes in training, validation and test sets.

\begin{table}
\caption[Class frequency in training, validation and test sets.]{Class frequency in training, validation and test sets. \REFEREE{Columns: AGN types in our dataset. Rows: subsets of the complete AGN spectra dataset used in supervised classification.} \label{tab:train_val_test_sets}
}
\centering
\begin{tabular}{llll}
\hline \hline
 & Type 1 & Type 2 & Int. \\ 
 \hline
Train & 441 & 1370 & 330\\
Validation & 107 & 352 & 76\\
Test & 132 & 423 & 115\\
\hline
\end{tabular}
\end{table}


We then performed a hyperparameter optimization for our SVM classifier using {a grid search approach}. The parameters optimized were {cost} $C$, that is the regularization parameter (the strength of the regularization is inversely proportional to $C$, {which represents the cost of misclassification for a soft-margin SVM}), and $\gamma$, {a kernel coefficient used only for Polynomial kernels or Radial Basis Function (RBF) kernels} that can be seen as the inverse of the radius of influence of samples selected by the model as support vectors. {The choice of the kernel used was also itself subject to optimization.}
The grid search optimization was firstly applied to a wide range for parameter $C$, going from $5 \times 10^{-4}$ to $5 \times 10^{4}$ on an equally-spaced logarithmic grid, and then interactively restricted around the best value until the F-score stopped improving (i.e. the fourth decimal digit remained constant). The range investigated for $\gamma$ was narrower, going from $0.005$ to $5.0$. Both ranges were selected while keeping in mind the trade-off between computational requirements and the ability to satisfactorily cover hyperparameter space.

The hyperparameters optimization was performed for four different kernels: linear, polynomial of degree 2 and 3, and RBF. The performance was evaluated with {the} F-score.
{F-score is defined as the harmonic mean of precision and recall \citep[][]{van1979information, chinchor1992statistical}, where precision is the number of true positives (TP) divided by the total number of samples classified as positive (that is TP plus false positive-- FP), and recall is the number of true positives divided by the number of all the actual positive samples (that is true positive plus false negatives --FN)}.

Based on this definitions, we can express precision, recall and F-score as follows:
\begin{align}
    P &= \frac{\mathrm{TP}}{\mathrm{TP} + \mathrm{FP}} \\
    R &= \frac{\mathrm{TP}}{\mathrm{TP} + \mathrm{FN}} \\
    F &= 2 \cdot \frac{P \cdot R}{P + R}.
\end{align}
In astronomical literature precision is often referred to as purity and recall as completeness.

A high value (close to 1) of the F-score means that the classifier is able to correctly classify most of the data, achieving both good precision and good recall.
{These definitions of course apply to a given class, where positive means a member of that class and negative a non-member. Their extension to a multi-class setting is straightforward by taking the mean over the different classes.}

It was found that the kernel with the highest performance, that is highest F-score, for multiclass classification was the linear one and the best regularization parameter was $C = 0.07$, while for two-class {classification all the kernels achieved nominally perfect results except for the polynomial of degree 3}. These metrics were calculated on the validation set. The performances of the four different models corresponding to the four kernels can be seen in Tab.~\ref{tab:gridsearch} for multiclass classification, and in Tab.~ \ref{tab:gridsearch_two} for two-class classification.

\begin{table}
\caption[F-score on validation set for four different models (multiclass classification).]{F-score on validation set for the multiclass classification problem for four different models corresponding to different SVM kernels (from top to bottom): linear, radial basis, polynomial of degree two, polynomial of degree three. \REFEREE{Columns: kernel (first column), hyperparameters optimized in the classification context (second and third columns) and F-score value (last column). Rows: SVM kernels.} \label{tab:gridsearch}}  
\centering
\begin{tabular}{llll}
\hline \hline
Kernel & Optimized C & Optimized $\gamma$ & F-score \\
\hline
Linear & 0.0700 & N$/$A & 0.920 \\
RBF & 34.000 & 0.003 & 0.912 \\
Poly 2 & 0.0005 & 0.500 & 0.916 \\
Poly 3 & 0.0005 & 0.050 & 0.918 \\
\hline
\end{tabular}
\end{table}

\begin{table}
\caption[F-score on validation set for four different models (two-class classification).]{F-score on validation set for four different models (two-class classification). \REFEREE{Columns and rows as in Tab.~ \ref{tab:gridsearch}.} \label{tab:gridsearch_two}}  
\centering
\begin{tabular}{llll}
\hline \hline
Kernel & Optimized C & Optimized $\gamma$ & F-score \\ 
\hline
Linear & 0.40000 & N/A & 1.000 \\
RBF & 45.0000 & 0.005 & 1.000 \\
Poly 2 & 0.00005 & 0.600 & 1.000 \\
Poly 3 & 0.00050 & 0.050 & 0.997 \\
\hline
\end{tabular}
\end{table}


We then trained the SVM using both the training and validation subsets and evaluated the model performance on the test set.
\subsection{Dependence on signal-to-noise ratio}
\begin{table}
\caption[{Performance metrics calculated on test set with incremental additive Gaussian noise.}]{{Performance metrics calculated on test set with incremental additive Gaussian noise. Noise standard deviation in units of flux value at 5100$\AA$ of the original spectrum. \REFEREE{Columns: Gaussian standard deviation noise values. Rows: Mean F-score (first row) and F-score values for every type in the dataset.}}\label{tab:results_noise}}  
\centering
\begin{tabular}{lllllllll}
\hline \hline
Noise $\sigma$ & 0.1 & 0.2 & 0.4 & 1.0 & 2.0 & 3.0 & 4.0 \\ 
\hline
Mean & 0.92 & 0.89 & 0.82 & 0.74 & 0.68 & 0.57 & 0.56 \\ 
Type 1 & 1.00 & 1.00 & 1.00 & 0.99 & 0.83 & 0.64 & 0.62 \\ 
Type 2 & 0.95 & 0.92 & 0.86 & 0.76 & 0.68 & 0.59 & 0.60 \\ 
Int.& 0.82 & 0.74 & 0.63 & 0.48 & 0.38 & 0.33 & 0.27 \\ 
\hline
\end{tabular}
\end{table}

\begin{figure*}[bth]
\subfloat[Noise factor values: 0.1, 0.2, 0.3, 0.4.]
{\label{fig:noise1}
\includegraphics[width=.50\linewidth]{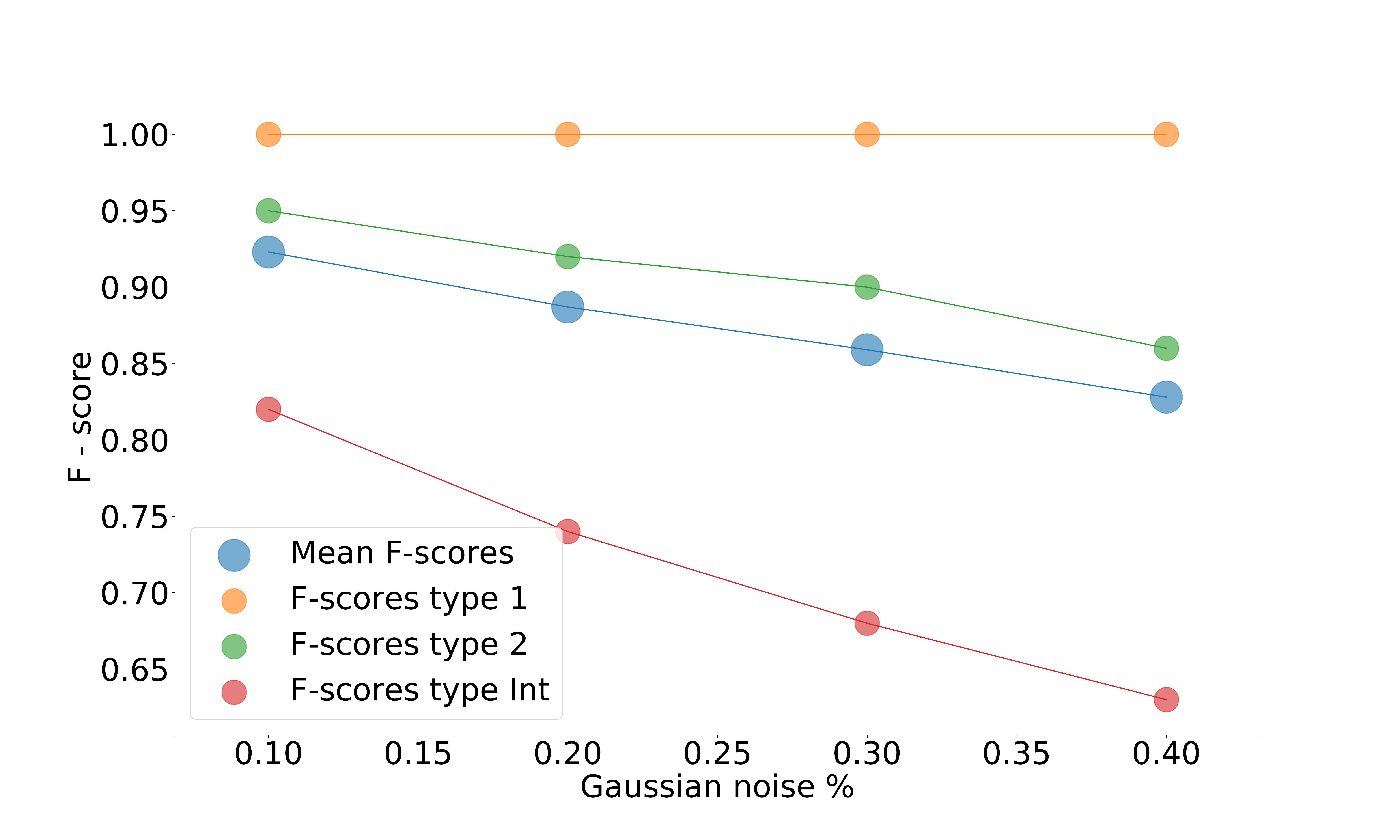}} \quad
\subfloat[Noise factor values: 1.0, 2.0, 3.0, 4.0.]
{\label{fig:noise2}
\includegraphics[width=.50\linewidth]{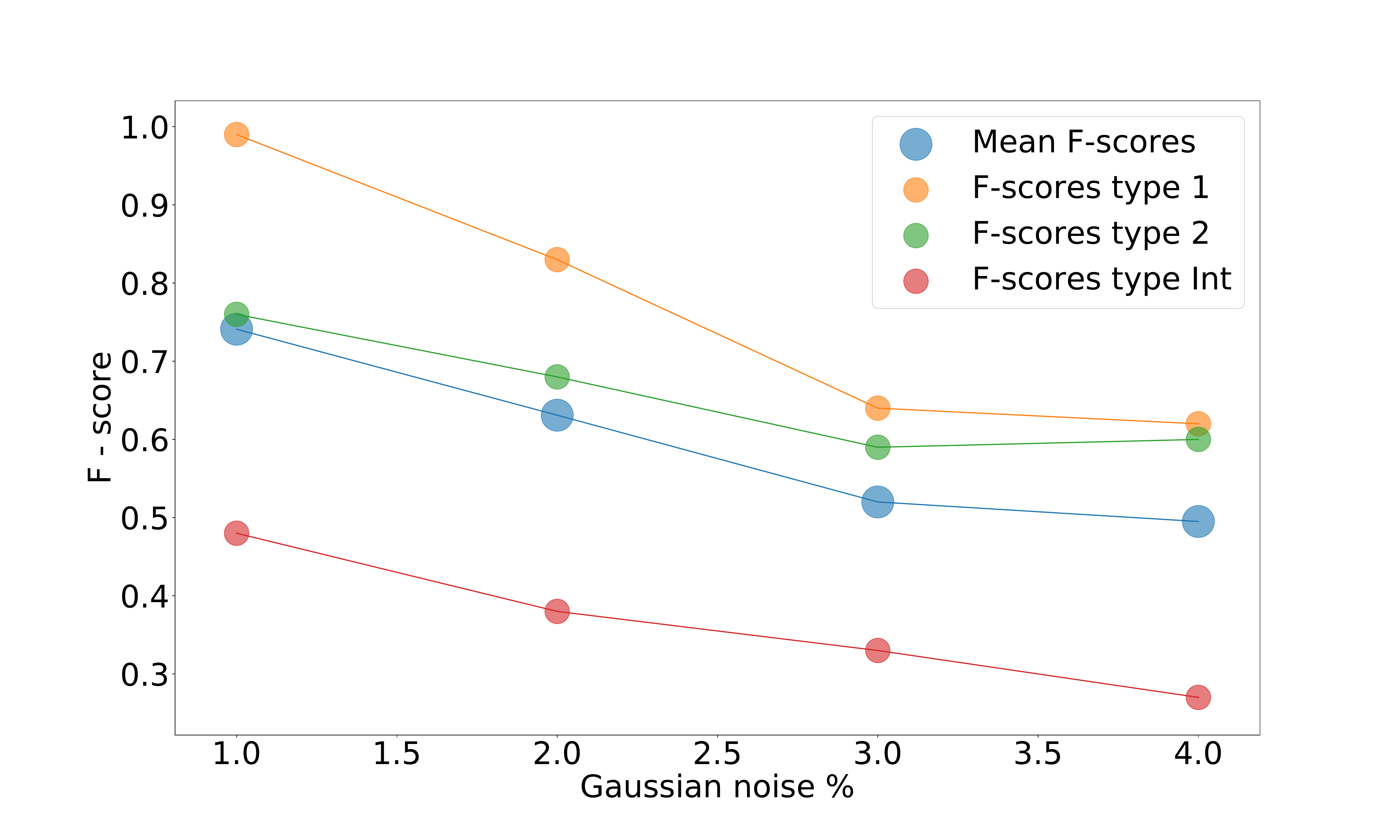}}
\caption[]{{F-scores for linear SVM on data with incremental noise. Blue: mean F-scores, orange: F-scores for type 1, green: F-scores for type 2 and red: F-scores for intermediate type.}}\label{fig:noises1}
\end{figure*}

{We explored how the performance of our SVM (trained on both the training and validation subsets) changes when we add Gaussian noise onto the test-set spectra. The noise's standard deviation was taken proportional to the flux value corresponding to 5100$\AA$, with the following values as proportionality factors: $0.1$, $0.2$, $0.3$, $0.4$, $1.0$, $2.0$, $3.0$, and $4.0$. The metrics can be seen in Tab.~\ref{tab:results_noise} and in Fig.~\ref{fig:noise1} and \ref{fig:noise2}.}
{As can be seen, for small values of the noise, the mean F-score remains above 0.8, but decreases almost linearly with increasing noise factor, while the Type 1 F-score initially remains equal to $1.0$. Type 2 F-score decreases similarly as the mean F-score, but remains above $0.85$. On the contrary the intermediate type F-score decreases rapidly with the noise factor, reaching $0.63$ for noise factor $0.4$.
With higher values for the noise factor, all the F-scores are below $0.8$, exception made for the first two values of type 1 F-score that remain above $0.8$ for noise factor $1.0$ and noise factor $2.0$. It is worth to notice that for higher values of the noise factor, type 1 F-score decreases rapidly, in contrast to the F-scores of both type 2 and intermediate. This can indicate that in general spectra characterized by a low S/N ratio are harder to classify and that the SVM classifier that we used begins to missclassify also type 1 AGNs for high values of the noise, but the confidence for type 2 and intermediate does not change considerably after some value of the noise factor.}

\section{Interpretability framework}
\subsection{Classifier gradient visualized as saliency map}
In order to gain insights on what our SVM classifiers have learned, we followed the \textit{saliency map} approach that found wide application in the context of deep neural networks \citep[][]{simonyan2013deep} and has proven very useful in the interpretation of image classifiers, showing which parts of a given image contribute the most to the image's predicted classification.
{In this paper we refer exclusively to the second meaning of the term saliency map defined in the  \cite{simonyan2013deep} paper, that is an image (in our case a one-dimensional array representing a given AGN spectrum) where each pixel represents the derivative of the class score with respect to the value of the corresponding pixel of a given image as per their Eq.~4.}

In the context of our work we built saliency maps as follows:
\begin{itemize}
    \item we considered the {class score or, loosely speaking, the probability} associated by our classifier to a given sample's predicted class, $p_c(f_i)$, where $f_i$ is the flux at wavelength $\lambda_i$ for a given spectrum and $c$ is the predicted class, for example, Type I;
    \item we then computed a numerical approximation to the gradient $g_i = \nabla_f \log p_c(f_i)$, which yields a vector the same length as the original spectrum;
    \item finally, we visualized the computed gradient vector as a color coding on top of the original spectrum, with blue (orange) corresponding to wavelengths for which the associated component of the gradient is positive (negative)
\end{itemize}

To compute the gradient $g_i$, each feature $f_{i}$ of a chosen sample is individually perturbed by a certain value $e_i$, forming \textit{n} spectra with the i-th feature perturbed, where \textit{n} is the number of features, that in this work is equal to $1000$. Then our SVM model is used to re-classify these perturbed spectra, obtaining a value of $p_c(f_j + \delta_{ij} e_i)$ for each one. Since the perturbation was chosen as $e_i = 0.01 f_i$ (i.e. a $1\%$ perturbation), $p_c(f_j + \delta_{ij} e_i) - p_c(f_i)$/$e_i$ approximates the $i-th$ component of the gradient of $\log p_c$.



A $g_i$ value close to zero (shown in white in the map) means that a perturbation of the i-th feature does not change the confidence of the classifier in classifying the spectrum as belonging to a specific class; a positive value (shown in blue in the map) means that a perturbation of the i-th feature strengthens the confidence of the {classifier's prediction for the given class (increases the class score)} and a negative value (shown in orange in the map) reduces it.

\subsection{Dimensionality reduction for visualization}
{The} t-distributed stochastic neighbor embedding \citep[t-SNE][]{maaten2008} is an unsupervised dimensionality reduction algorithm used for visualization and data exploration in many machine learning settings. The goal of dimensionality reduction is to map high-dimensional data to a lower-dimensional space (in our case the plane) while preserving the pairwise distances of points. This is impossible to do rigorously, because the high-dimensional space cannot be embedded in the plane, but t-SNE achieves this approximately by prioritizing the distances of points that are near to each other, so short distances are distorted the least, while the large-scale structure of the data-set is mostly lost.
{This is obtained by minimizing a loss}
\begin{equation}
\label{ciccia}
    L = - \sum_{i \neq j} p_{ij} \log{q_{ij}/p_{ij}} 
\end{equation}
{where $p_{ij}$ is a similarity measure between points $i$ and $j$ in the original high dimensional space and $q_{ij}$ is a (different) similarity measure in the low dimensional space. While $p_{ij}$ decays as a Gaussian with the distance between point $i$ and $j$, $q_{ij}$ decays like a Student's t-distribution with one degree of freedom, hence the name of the algorithm.
We can see from Eq.~\ref{ciccia} that points that are far from each other in the high dimensional space do not contribute much to the loss, as their $p_{ij}$ goes to zero exponentially with squared distance.
The outcome of t-SNE depends on the perplexity hyperparameter, which drives the standard deviation of the Gaussian used to define $p_{ij}$ and can be loosely interpreted as the typical size of the subgroups expected in a given dataset. A practical illustration of the effect of varying perplexity can be found in \cite{wattenberg2016how}.
Since perplexity can be set at the discretion of the user of t-SNE, results that depend strongly on this parameter, such as e.g. clustering structure that shows up only for a narrow range of values of perplexity, should not be blindly trusted. In the following we make sure to  test a wide range of perplexity values.
We use t-SNE in the scikit-learn implementation \citep[][]{scikit-learn} for Python. While our main use for t-SNE visualization is to show where the AGN spectra we selected for inspection through saliency maps are located, which is particularly useful for misclassified spectra, we will also gain some useful insight on the structure of our dataset through this approach, as shown below.}

\section{Results}
\subsection{Classifier performance}

\begin{table}
\caption[Metrics values on test set for linear, RBF and polynomial kernels for multiclass classification.]{Metrics values on test set for linear, RBF and polynomial kernels for multiclass classification. \REFEREE{Columns: values of precision, recall and F-score. Rows: SVM kernels.} \label{tab:results_metrics_multi}}  
\centering
\begin{tabular}{llll}
\hline \hline
Kernel & Precision & Recall & F-score \\ 
\hline
Linear & 0.948 & 0.936 & 0.942 \\
RBF & 0.945 & 0.927 & 0.935 \\
Poly 2 & 0.925 & 0.928 & 0.927 \\
Poly 3 & 0.935 & 0.932 & 0.933 \\
\hline
\end{tabular}
\end{table}

\begin{table}
\caption[Metrics values on test set for linear, RBF and polynomial kernels for two-class classification.]{Metrics values on test set for linear, RBF and polynomial kernels for two-class classification. \REFEREE{Columns and rows as in Tab.~ \ref{tab:results_metrics_multi}.} \label{tab:results_metrics_two}}  
\centering
\begin{tabular}{llll}
\hline \hline
Kernel & Precision & Recall & F-score \\ 
\hline
Linear & 1.0 & 1.0 & 1.0 \\
RBF & 1.0 & 1.0 & 1.0 \\
Poly 2 & 1.0 & 1.0 & 1.0 \\
Poly 3 & 1.0 & 1.0 & 1.0 \\
\hline
\end{tabular}
\end{table}

\begin{table}
\caption[Precision, recall and F-score obtained by SVM with linear kernel for every class in the test set.]{Precision, recall and F-score obtained by SVM with linear kernel for every class in the test set. \REFEREE{Columns as in Tab.~ \ref{tab:results_metrics_multi}. Rows: AGN types.} \label{tab:results_metrics_multi_allclasses}}  
\centering
\begin{tabular}{llll}
\hline \hline
Type & Precision & Recall & F-score \\ 
\hline
1 & 1.0 & 1.0 & 1.0 \\
2 & 0.96 & 0.97 & 0.96 \\
Int. & 0.89 & 0.83 & 0.86 \\
\hline
\end{tabular}
\end{table}

\begin{figure*}[bth]
\subfloat[Confusion matrix for RBF kernel.]
{\label{fig:heat_rbf_test}
\includegraphics[width=.40\linewidth]{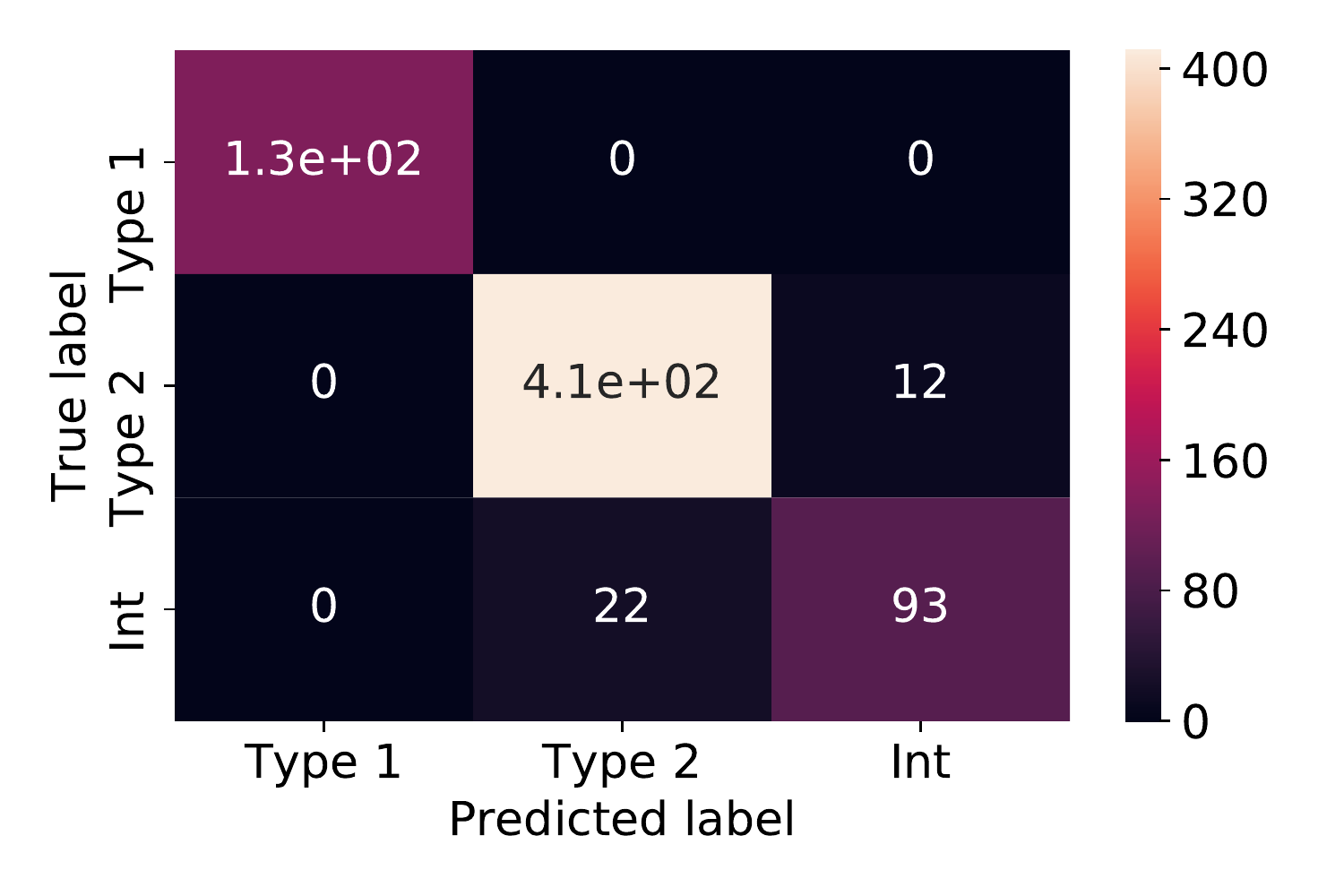}} \quad
\subfloat[Confusion matrix for linear kernel.]
{\label{fig:heat_lin_test}
\includegraphics[width=.40\linewidth]{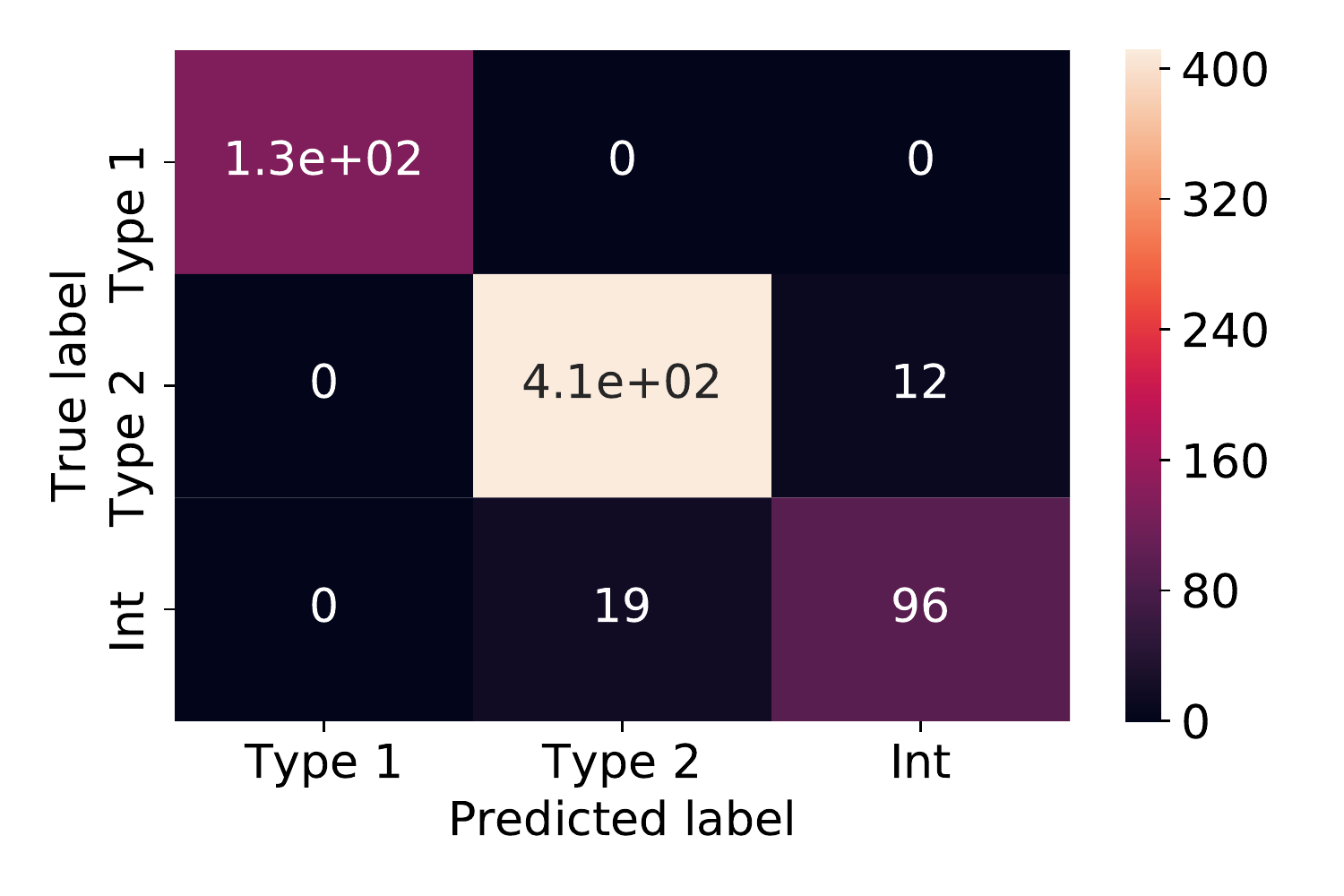}}
\caption[Confusion matrix heatmaps for SVM classification over test set.]{Confusion matrix heatmaps for SVM classification over test set. Horizontal axis: labels predicted by the classifier. Vertical axis: true labels for the samples.}\label{fig:heatmaps_test_lin_rbf}
\end{figure*}

\begin{figure*}[bth]
\subfloat[Normalized confusion matrix for RBF kernel.]
{\label{fig:norm_heat_rbf_test}
\includegraphics[width=.40\linewidth]{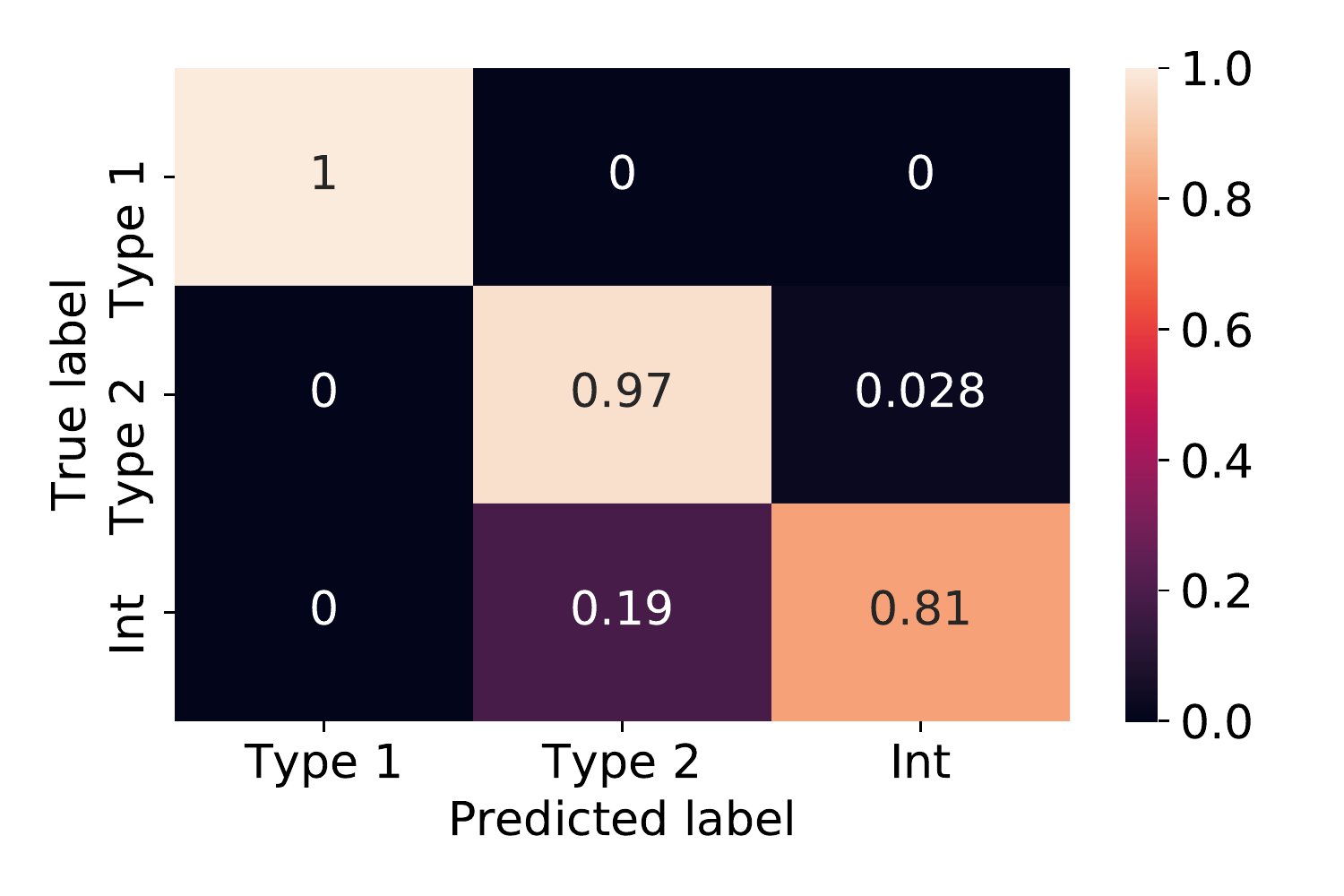}} \quad
\subfloat[Normalized confusion matrix for linear kernel.]
{\label{fig:norm_heat_lin_test}
\includegraphics[width=.40\linewidth]{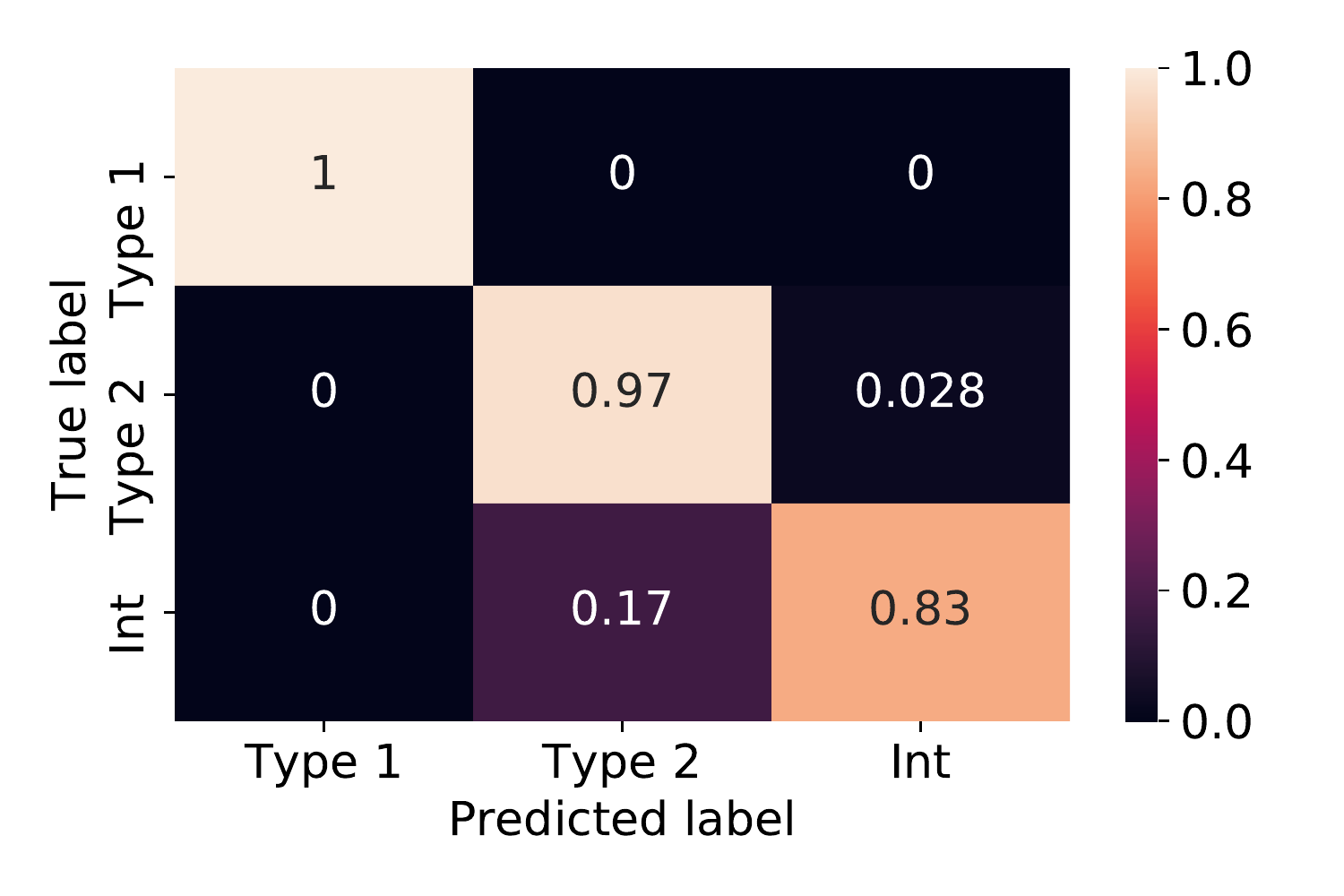}}
\caption[Normalized confusion matrix heatmaps for SVM classification over test set.]{Normalized confusion matrix heatmaps for SVM classification over test set. Horizontal axis: labels predicted by the classifier. Vertical axis: true labels for the samples.}\label{fig:norm_heatmaps_test_lin_rbf}
\end{figure*}


The metrics values reached by our models on our test set are reported in Tab.~\ref{tab:results_metrics_multi} for multiclass classification and in Table \ref{tab:results_metrics_two} for two-class classification. We find that they are comparable to those obtained on the validation set, suggesting no overfitting is occurring. In Table \ref{tab:results_metrics_multi_allclasses} can also be seen the value of precision, recall and F-score obtained by our best model for the three classes: type 1, type 2 and intermediate type AGN.
It is clear that separating type 1 and type 2 can be easily done by every kernel, with the right choice of hyperparameters. On the other hand, the multiclass classification including intermediate type spectra, turns out to be a more difficult task to solve, requiring a careful choice of hyperparameters in order to achieve high performances.

The confusion matrix for the two best models for multiclass classification can be seen in Fig. \ref{fig:heatmaps_test_lin_rbf} and the normalized confusion matrix in Fig. \ref{fig:heatmaps_test_lin_rbf}. Even if the linear kernel performs slightly better than the RBF, both models are able to classify the majority of the spectra, failing only in the classification of a small number of type 2 and intermediate type spectra. Specifically 22 intermediate spectra, over 115 total, were classified as type 2 by the RBF model and 12 type 2, over 423 total, as intermediate; the linear model failed to classify 19 intermediate spectra, over the same 115 total, by classifying them as type 2 and 12 type 2, over the same 423 total, classified as intermediate. To a point, this is an expected result, because the distinction between type 2 and intermediate AGN is difficult in the presence of spectra with low S/N. Therefore, this uncertainty in the distinction between intermediate and type 2 AGN spectra in presence of a low S/N, can affect the automated classification result. 

\subsection{Training time complexity}

\begin{table}
\caption[Training computational time for various kernels. Every time measurement is calculated by taking the average of 10 trainings for every kernel. All measures are in second.]{Training computational time for various kernels. Every time measurement is calculated by taking the average of 10 training for every kernel. All measures are in seconds. \REFEREE{Columns: SVM kernels. Rows: computational times.} \label{tab:time_compl}}  
\centering
\begin{tabular}{llll}
\hline \hline
Linear & RBF & Poly 2 & Poly 3 \\ 
\hline
15.13 s & 13.34 s & 12.94 s & 12.96 s\\
\hline
\end{tabular}
\end{table}

Every kernel was also evaluated in terms of the computational time of the training. The computational time is evaluated by taking the time average of 10 different trainings (using both train and validation sets for this purpose) for every kernel. \REFEREE{The results presented in Tab.~\ref{tab:time_compl} show that the two polynomial kernels are the fastest, in particular the polynomial of degree 2. Surprisingly the linear kernel appears to be the slowest.} A possible explanation could be the fact that the scikit-learn implementation used in this work (libsvm-based \citep[][]{libsvm}) is less efficient for the linear case, as stated in the scikit-learn documentation \citep[][]{scikit-learn}.
The documentation also provides an estimation of the time complexity, in big O notation, of the SVM implementation, that scales between $O(n_{features} \: x \: n_{samples}^{2})$ and $O(n_{features} \: x \: n_{samples}^{3})$ \citep[][]{scikit-learn}.
Every computation in this step was performed on a Intel(R) Core(TM) $i7-6700HQ$ CPU (2.60$GHz$).

\subsection{Visualizing spectra with t-SNE}

Thanks to the interpolation used in this work, the AGN spectra space turns out to be $1000$-dimensional, while the original spectra comprised a variable number of points typically of order a few thousands. The dimensionality of feature space is still, however, quite high.
We then used t-SNE to map our spectra dataset to a plane. The algorithm was firstly applied to data not scaled and not mean normalized to compare its results between this case and the case with features pre-processed as described below.

\begin{figure}
\includegraphics[scale=0.2]{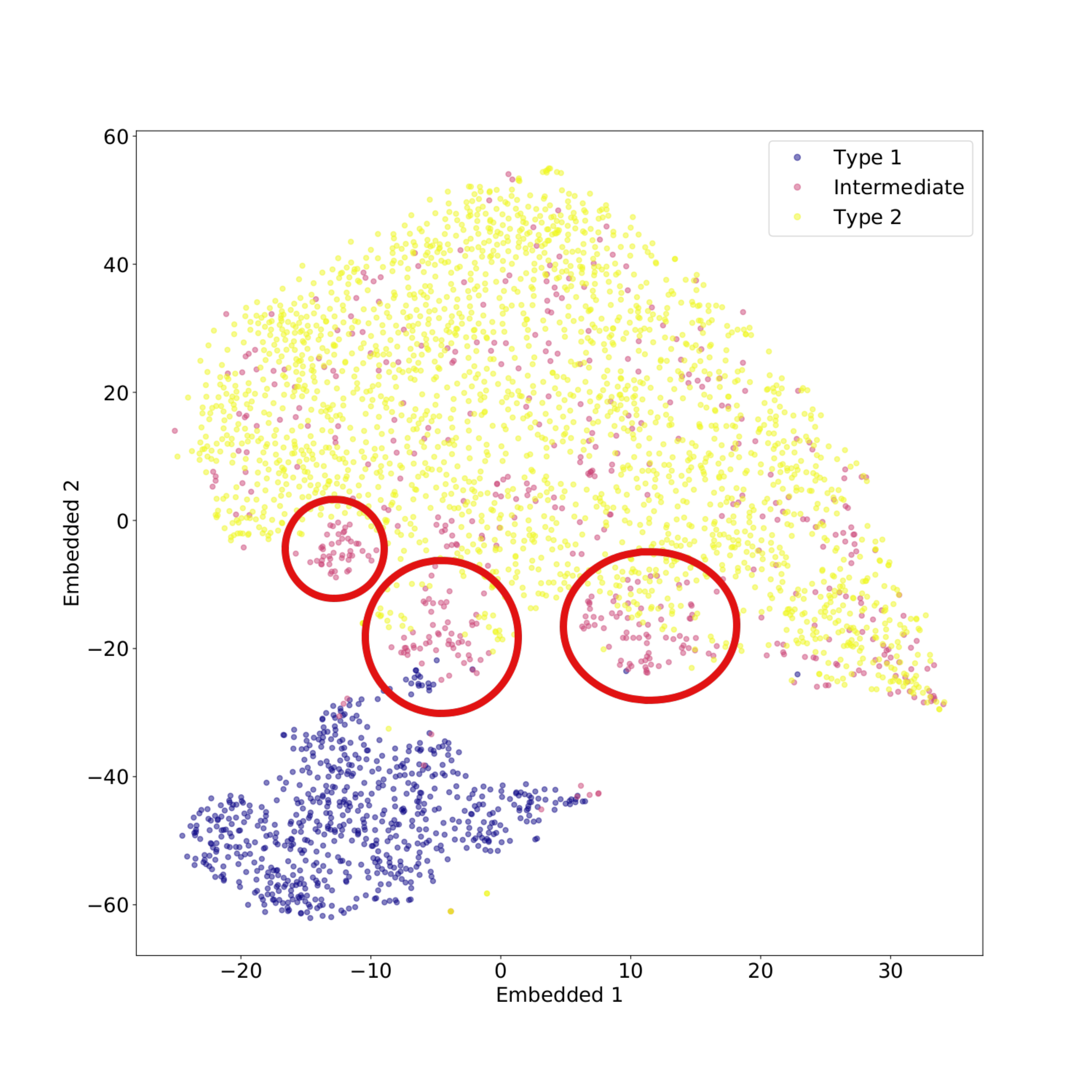}
\caption{t-SNE embedded plane for type 1, type 2 and intermediate type AGN. Perplexity: 50. \REFEREE{Blue points: type 1. Yellow points: type 2. Red points: intermediate type. Red circles: possible intermediate subgroups identified by the t-SNE algorithm.}}
\label{fig:tsne_tot}
\end{figure}

The result of t-SNE applied only to type 1 and type 2 AGN can be seen in Fig.~\ref{fig:tsne_1_2}. In the embedded plane, type 1 AGN and type 2 AGN are clearly separated, with just a few outliers. The perplexity parameter was set to $50$ in Fig.~\ref{fig:tsne_1_2}. With lower perplexities the separation between the two types was somewhat less clear, but still persisted as can be seen in Figure \ref{fig:t_sne_perplexities}. Additionally, some smaller-scale structures
can be appreciated. 
We also applied t-SNE to the whole dataset, including also intermediate type AGN. The result can be seen in Figure \ref{fig:tsne_tot} (in this case as well the perplexity was set to 50). Predictably, intermediate type AGN cannot be well separated from the other two classes, in particular from type 2 spectra with which they appear somewhat mixed. However, there is a clear cluster of intermediate type spectra connecting the regions occupied by type 1 and type 2 spectra, true to the definition of 'intermediate type'. While spurious groups may sometimes appear in t-SNE plots, this is likely a physically motivated feature, {since it persists while varying perplexity (see below).}
At the moment we can only speculate as to the physical meaning of the other two subclusters of intermediate type spectra which appear to gather in distinct islands at the extremes of the region occupied by type 2 spectra. {Perhaps this should be addressed by direct visual inspection of the spectra, as part of a future work.}
In Figure \ref{fig:t_sne_perplexities_whole} we plot the embedded spaces for various values of perplexity, showing that the main results we outlined here are robust for changes of the perplexity parameter; this is further discussed in appendix.

\subsection{Saliency maps}
\label{subsec:sm}

\begin{figure}
    \centering
    \includegraphics[width=0.95\columnwidth]{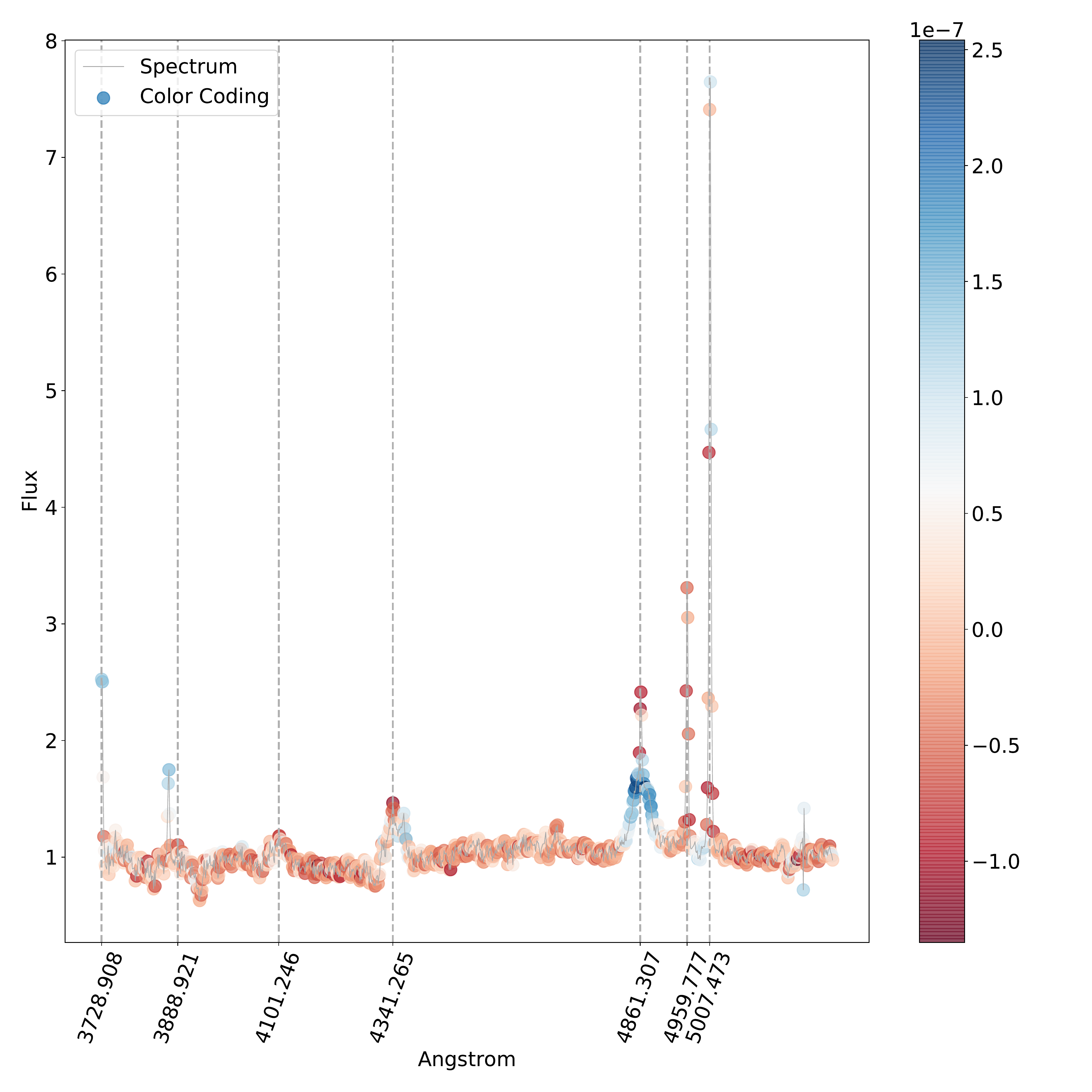}
    \caption{Intermediate-type spectrum that was correctly classified by our SVM. The regions of the spectrum shown in blue are those that most contributed towards its classification as an intermediate type, whereas those shown in red would reduce the SVM classification confidence if their flux were to increase. Several regions surrounding lines conventionally used for classification appear in blue, suggesting that our SVM model relied, in this case, on clues similar to those used by human experts. Notice in particular how the center of the $H \beta$ line appears in red, while the tails in blue: this corresponds to the classifier using the width of the $H \beta$ line for its decision.\label{fig:first_salmap}}
\end{figure}

\begin{figure}
    \centering
    \includegraphics[width=0.95\columnwidth]{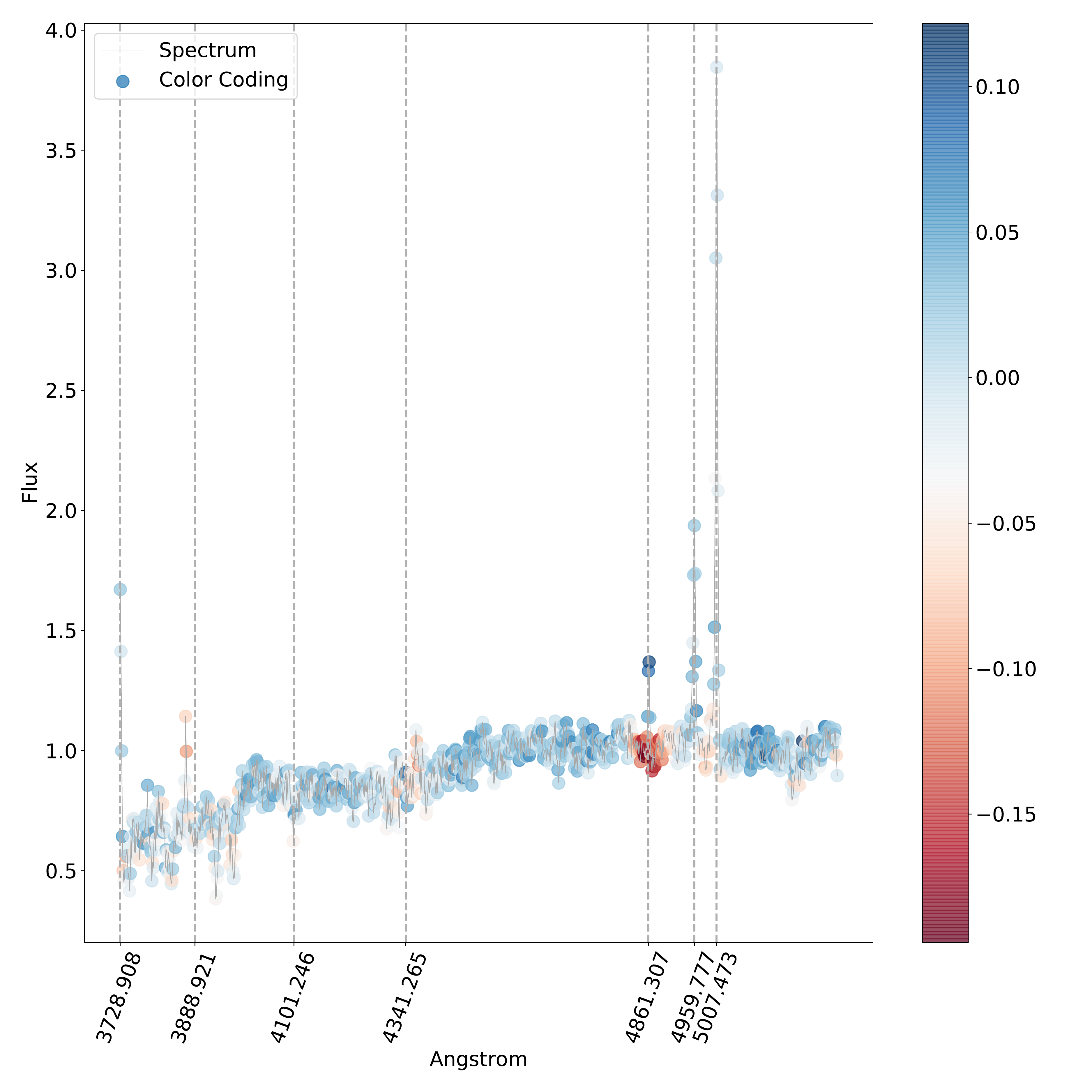}
    \caption{Intermediate type spectrum missclassified as type 2 with $76\%$ confidence. \REFEREE{The regions of the spectrum shown in blue are those that most contributed towards its classification as a type 2, whereas those shown in red would reduce the SVM classification confidence if their flux were to increase.}\label{fig:second_salmap}}
\end{figure}
In the following we consider directly the multiclass problem (Type 1, Type 2, Intermediate) because of its higher scientific interest and due to the fact that in the two-class problem we find no misclassified instances, i.e. we have a nominally perfect accuracy as discussed above. We thus used saliency maps to investigate missclassified and correctly classified spectra in the multiclass problem. The saliency maps for a correctly classified (Fig.~\ref{fig:first_salmap}) and a missclassified (Fig.~\ref{fig:second_salmap}) spectrum, show that the main optical lines used by humans to classify AGN spectra are also recognized by the SVM as important features.
{In every saliency map we plot the main lines that can be used to classify AGN spectra (namely $[O II] 3727$, $He I 3889$, $H \delta 4101$, $H \gamma 4340$, $H \beta 4861$, $[O III] 4959$ and $[O III] 5007$) are marked by grey vertical dashed lines.}
Especially the region around the $H \beta$ line plays an important role, as can be seen in both Fig.~\ref{fig:first_salmap} and Fig.~\ref{fig:second_salmap} where the center of the line and its tails appear in opposite colors, signifying the opposite effect on the class score of an increase in flux. In particular Fig.~\ref{fig:second_salmap} is an intermediate type confidently ($76\%$) misclassified as Type 2, with the model's decision depending mostly on the $H \beta$ line. This is apparent by looking at the region next to the $H \beta$ line, where the continuum is the reddest spot in the saliency map. This shows that the misclassification is largely due to the absence of a broad component in $H \beta$ (we recall here that red means that increasing the flux value at that location would reduce the confidence in the predicted class). Other regions of the saliency map that appear to contribute to the misclassification are the other hydrogen lines, but their contribution is very minor as evidenced by the color coding. Still, their pattern of color coding is similar, suggesting that lack of a broad component is the main driving feature for misclassification here. To properly classify this spectrum we likely would need to observe the $H \alpha$ line, which is not included in the current spectral range, otherwise an intermediate type $1.9$, which would show a broadening only on the $H \alpha$ line may appear as a Type 2, having a virtually unbroadened $H \beta$ line.
These findings should be contrasted with Fig.~\ref{fig:first_salmap}, where the color coding shows the same behaviour, but in reverse: there the tails of the $H \beta$ appear colored in blue, showing that increasing the flux there would lead to an even more confident classification as intermediate type. This applies similarly to the other hydrogen lines.

{For example increasing the flux in the tails of the $H \beta$ line, i.e. increasing its width for a given height of the central peak reduces} the classification probability of classifying {the spectrum in Fig.~\ref{fig:first_salmap} as intermediate, while it increases} the probability of classifying {it} as type 1, as one would expect; {increasing the flux in the center has exactly the opposite effect.}
This result is expected because a broader $H \beta$ profile, indicative of a type 1 AGN, would significantly change the shape of the spectrum next to the line.
{Interestingly, the saliency maps show that} the continuum between the $H \gamma$ and $H \beta$ {also affects the classification results}. {This is also reasonable, considering how the continuum differs between type 1 spectra and type 2 $/$ intermediate.} 

\begin{figure}
    \centering
    \includegraphics[width=0.95\columnwidth]{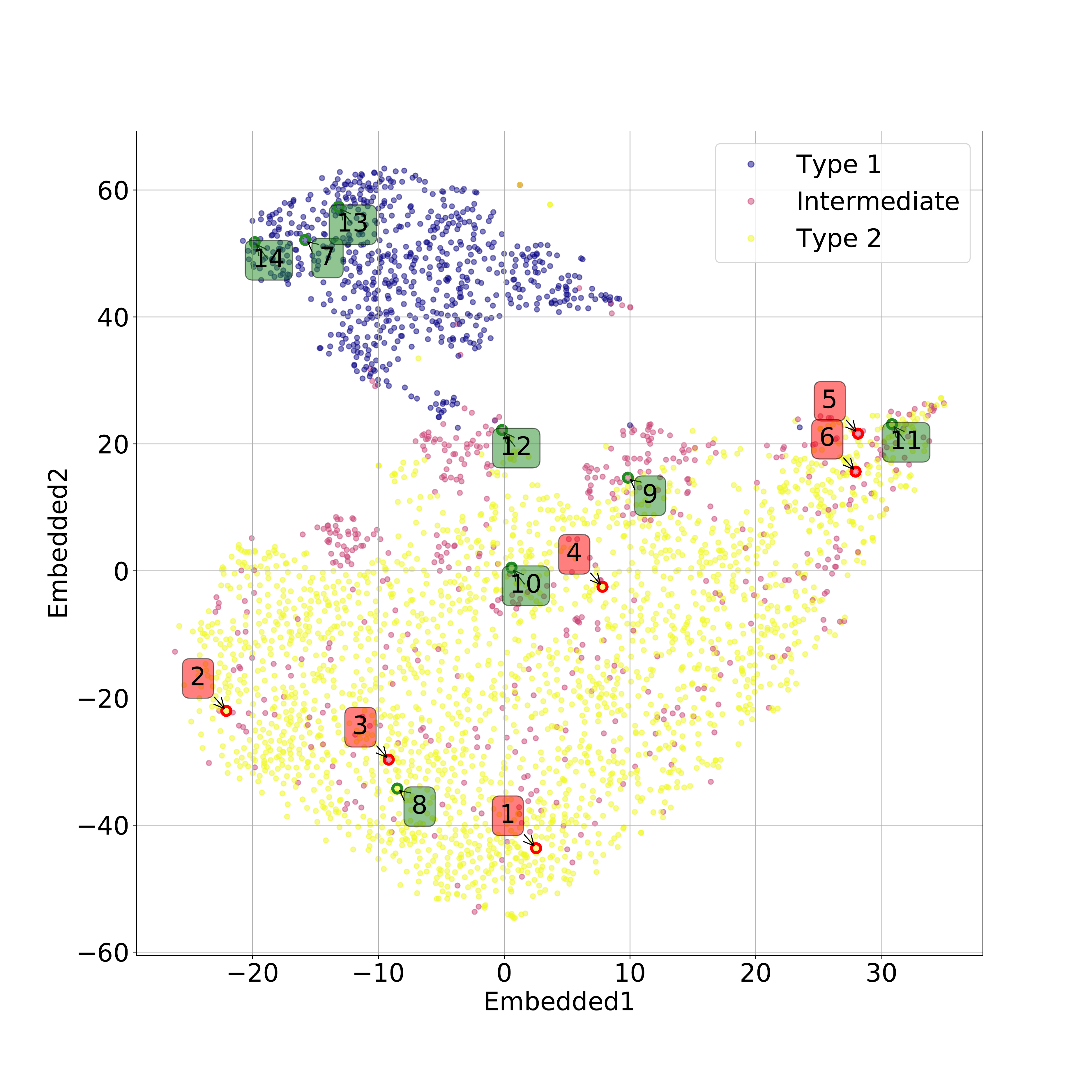}
    \caption[Selected spectra for which we computed and visualized saliency maps, shown in the t-SNE embedded plane.]{Randomly selected spectra for which we computed and visualized saliency maps, shown in the t-SNE embedded plane. Spectra that were correctly classified by our SVM are shown in green, while misclassified spectra are shown in red. \REFEREE{Color points as in Fig. \ref{fig:tsne_tot}. Green rectangles: correctly classified spectra for which saliency maps were calculated. Red rectangles: missclassified spectra for which saliency maps were calculated.}}
    \label{fig:tsneannot}
\end{figure}

\begin{figure*}[bth]
\renewcommand*\thesubfigure{\arabic{subfigure}} 
\subfloat[Type 2 missclassified as Int.]
{\label{fig:1miss}
\includegraphics[width=.3\linewidth]{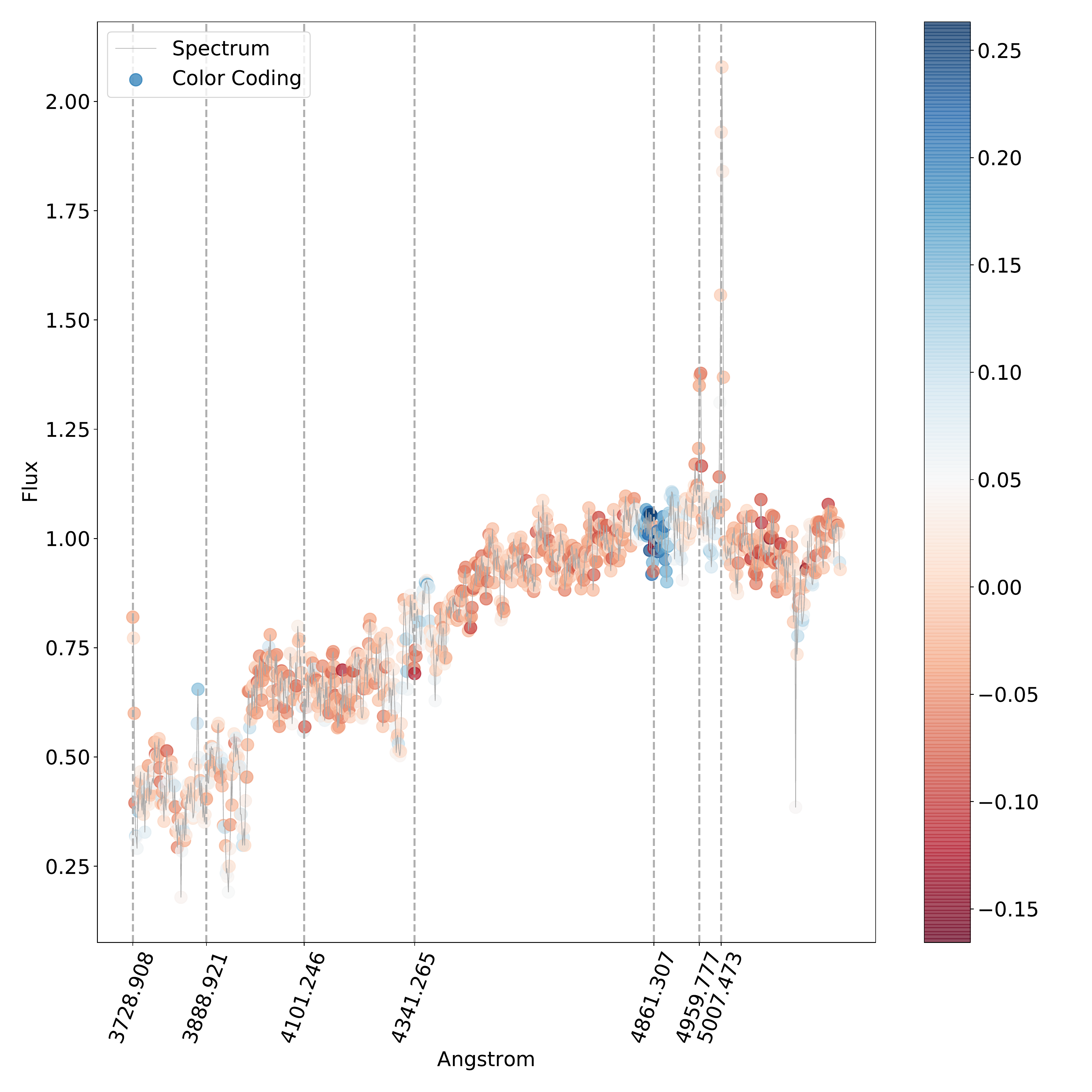}} \quad
\subfloat[Type 2 missclassified as Int.]
{\label{fig:2miss}
\includegraphics[width=.3\linewidth]{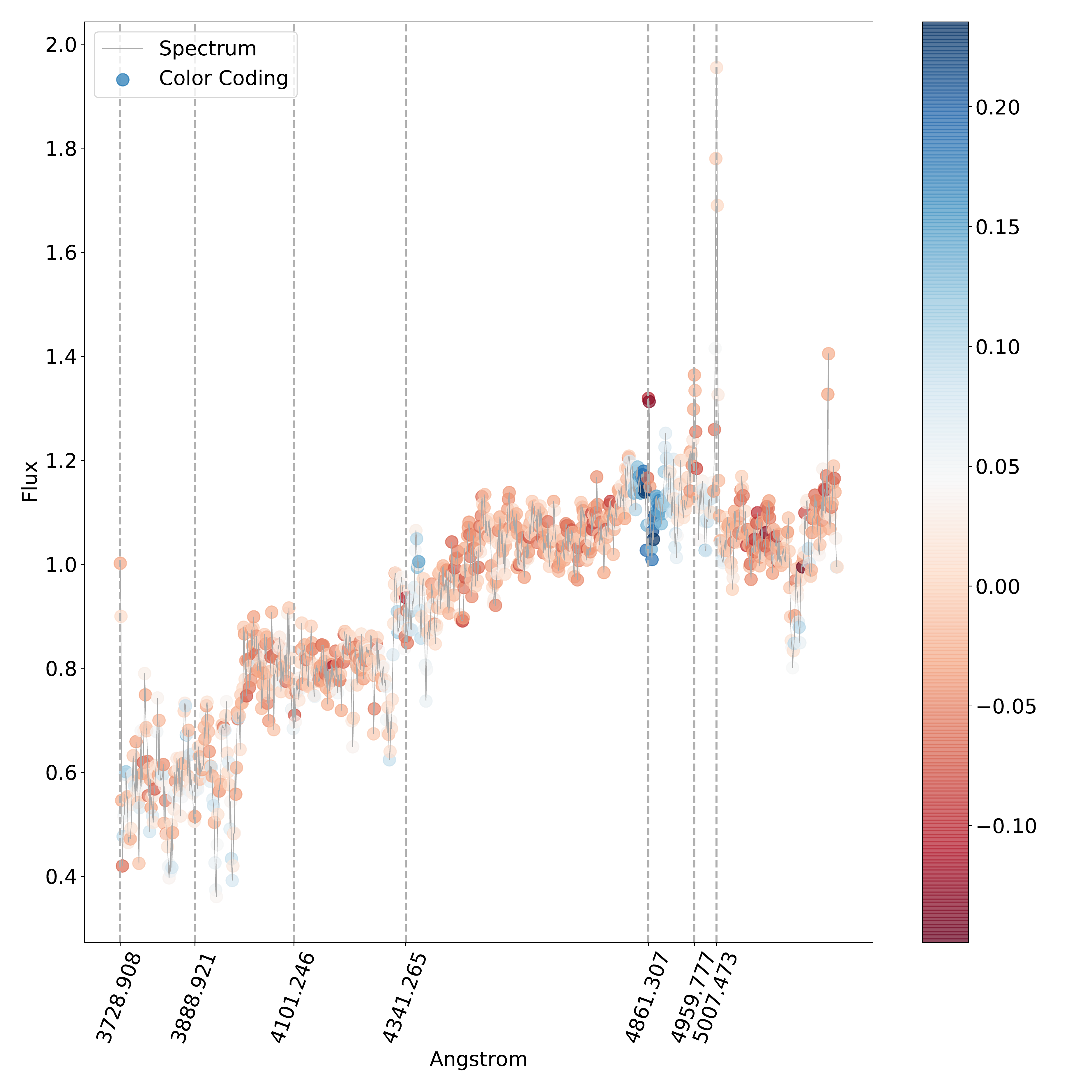}} \quad
\subfloat[Type 2 missclassified as Int.]
{\label{fig:4miss}
\includegraphics[width=.3\linewidth]{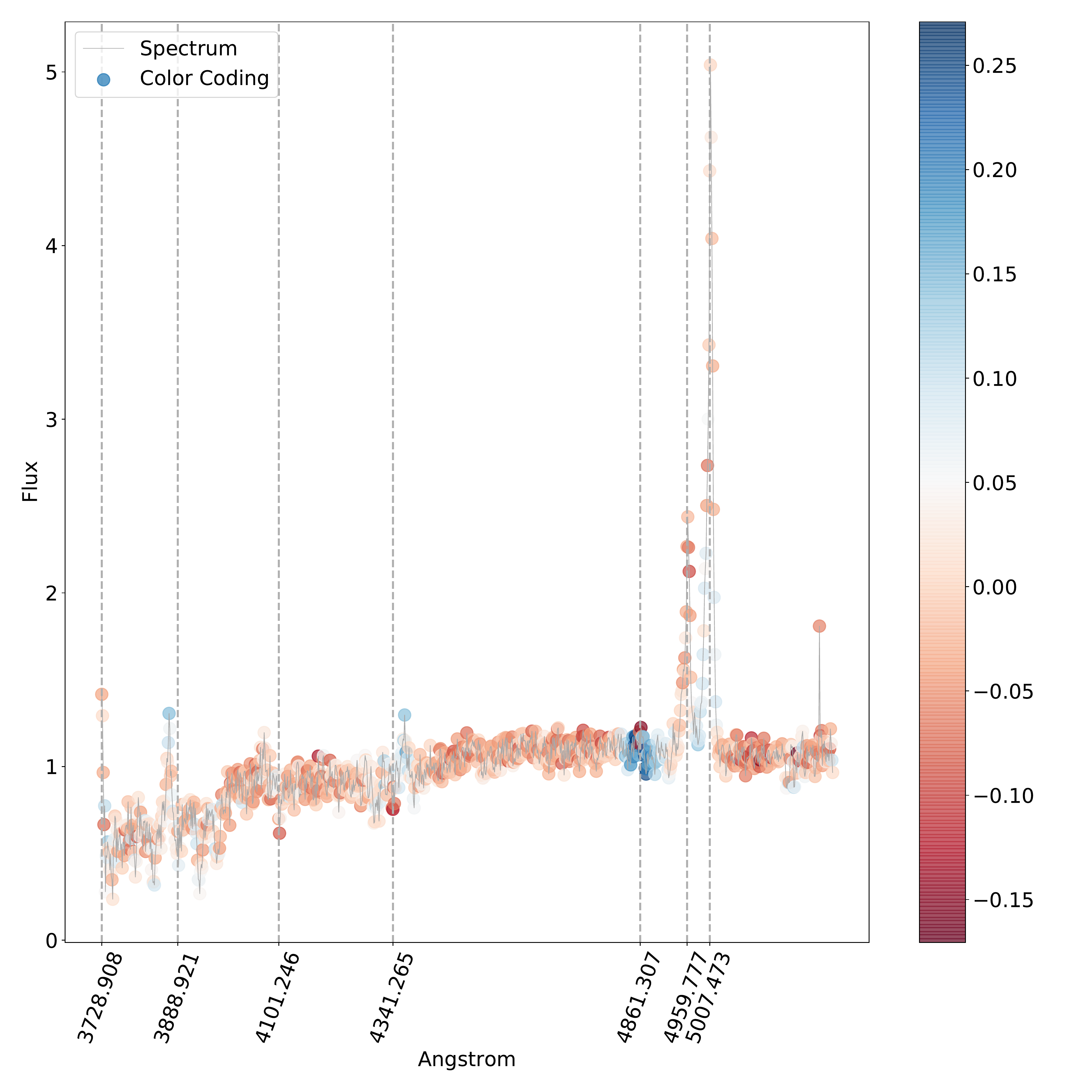}} \quad
\subfloat[Int. missclassified as type 2]
{\label{fig:5miss}
\includegraphics[width=.3\linewidth]{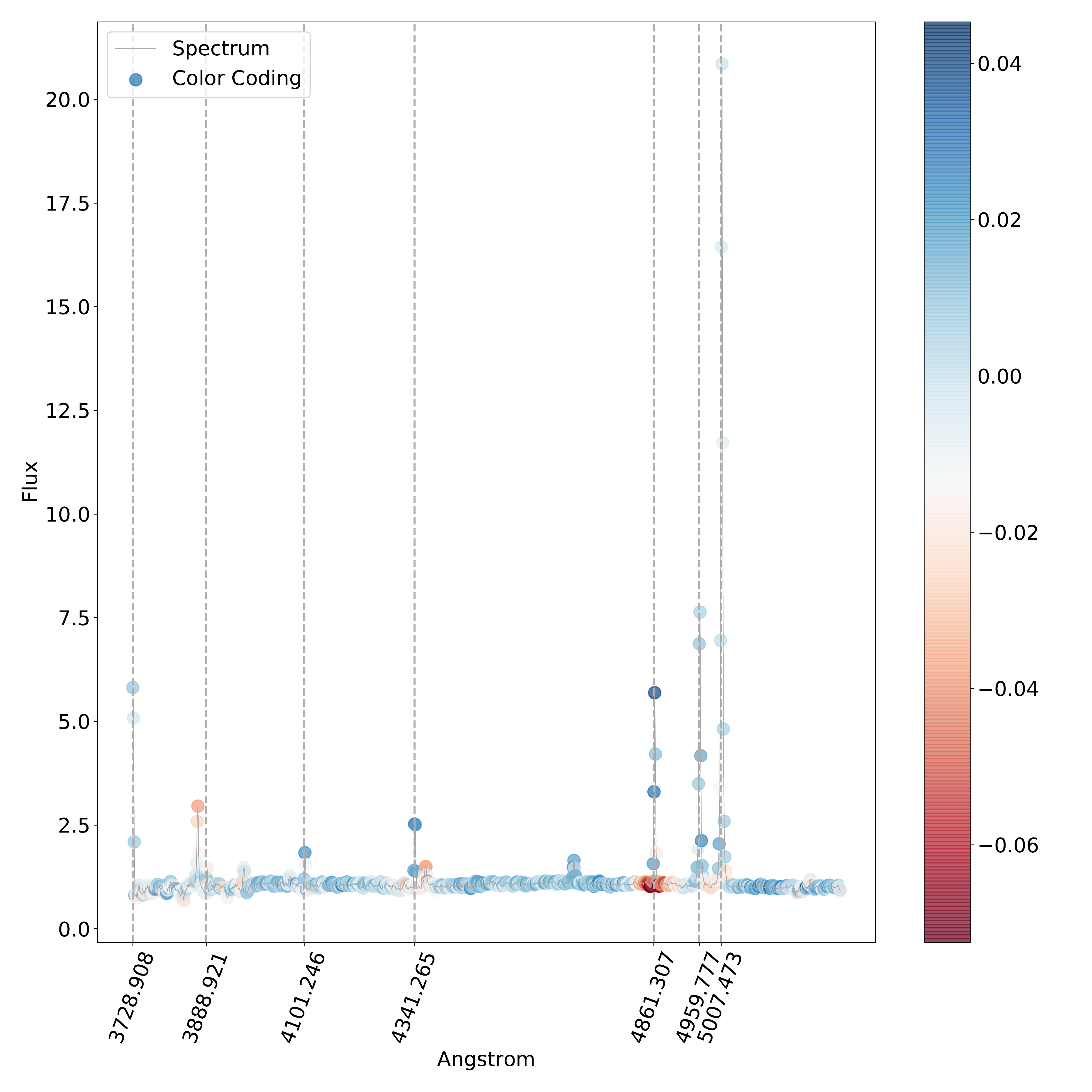}} \quad
\subfloat[Int. missclassified as type 2]
{\label{fig:6miss}
\includegraphics[width=.3\linewidth]{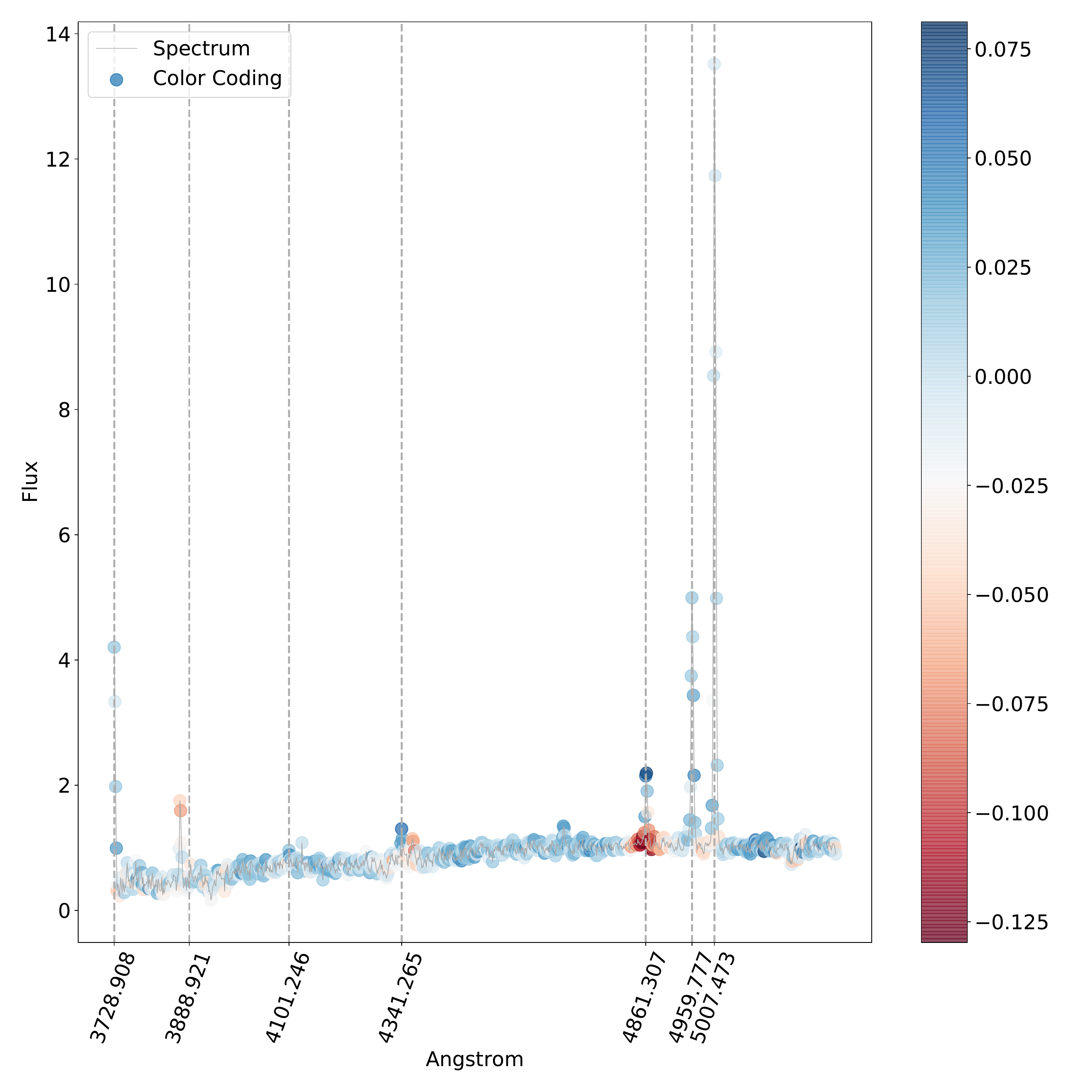}} \quad
\subfloat[Int. missclassified as type 2]
{\label{fig:3miss}
\includegraphics[width=.3\linewidth]{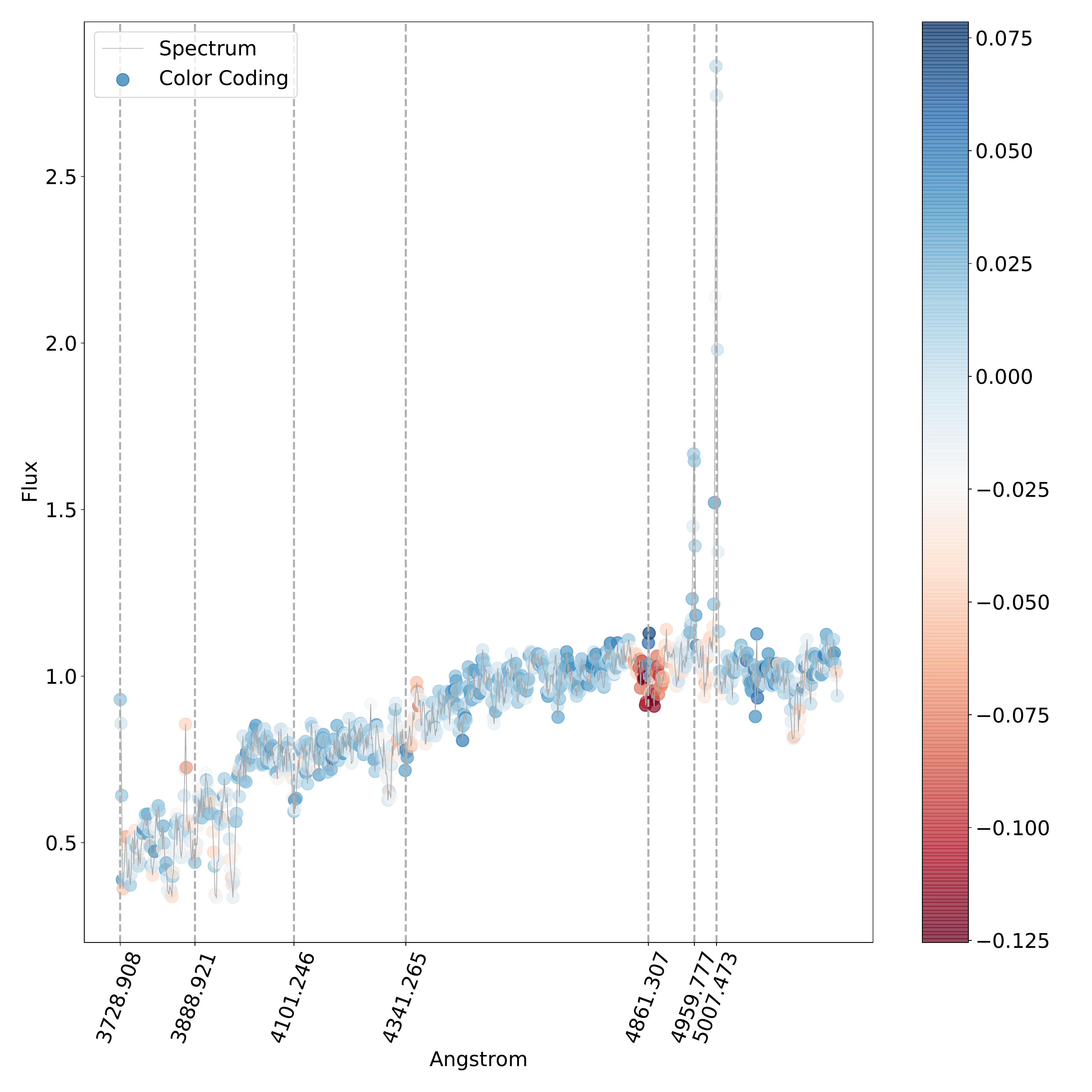}}
\caption[Saliency maps for missclassified spectra.]{Saliency maps for missclassified spectra. \REFEREE{The regions of the spectrum shown in blue are those that most contributed towards the classification chosen by our model, whereas those shown in red would reduce the SVM classification confidence if their flux were to increase.}}\label{fig:sal_maps_miss}
\end{figure*}

\begin{figure*}[bth]
\renewcommand*\thesubfigure{\arabic{subfigure}}
\addtocounter{subfigure}{6}
\subfloat[Type 1, correctly classified]
{\label{fig:7corr}
\includegraphics[width=.3\linewidth]{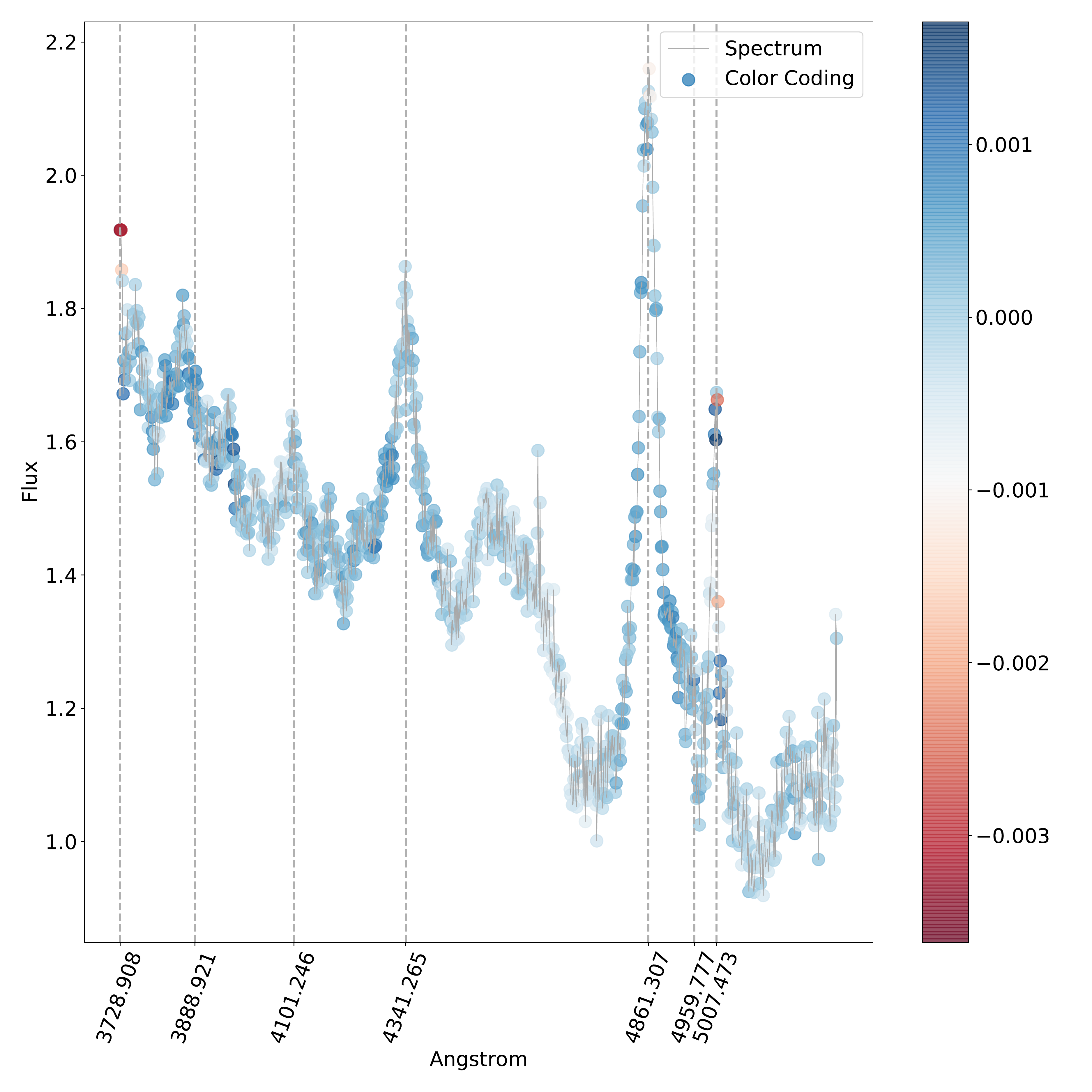}} \quad
\subfloat[Type 2, correctly classified]
{\label{fig:8corr}
\includegraphics[width=.3\linewidth]{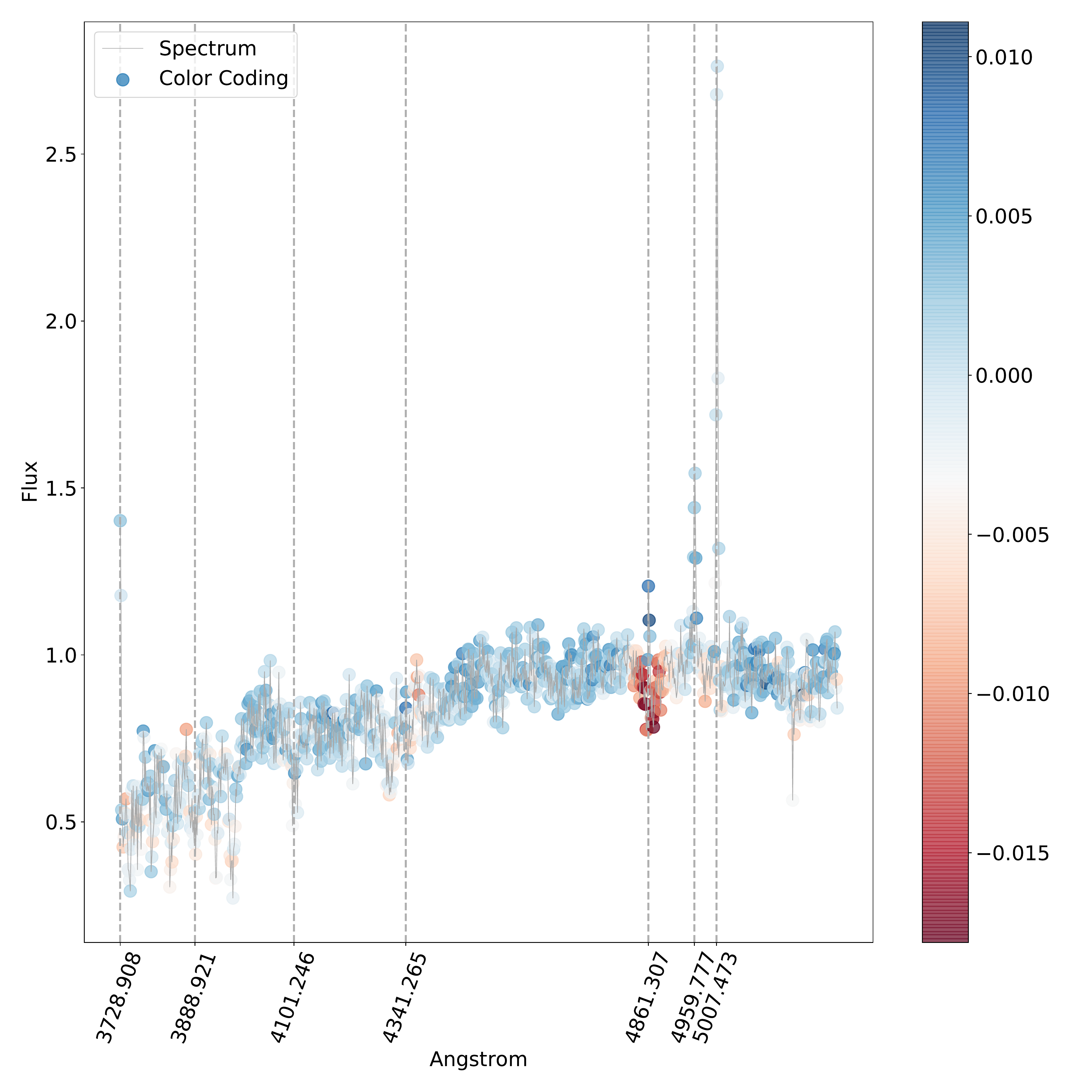}} \quad
\subfloat[Int. type, correctly classified]
{\label{fig:10corr}
\includegraphics[width=.3\linewidth]{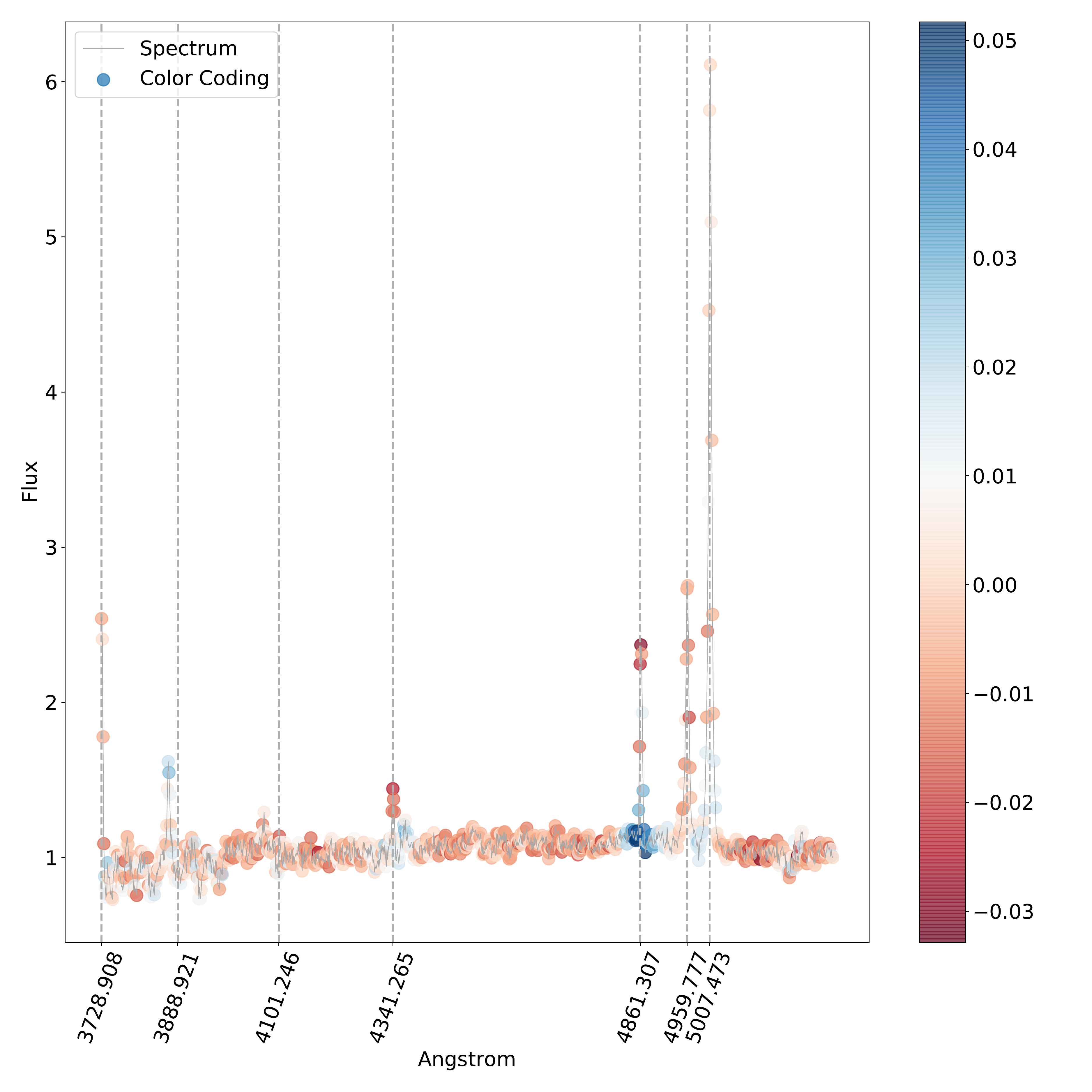}} \quad
\subfloat[Type 2, correctly classified]
{\label{fig:11corr}
\includegraphics[width=0.3\linewidth]{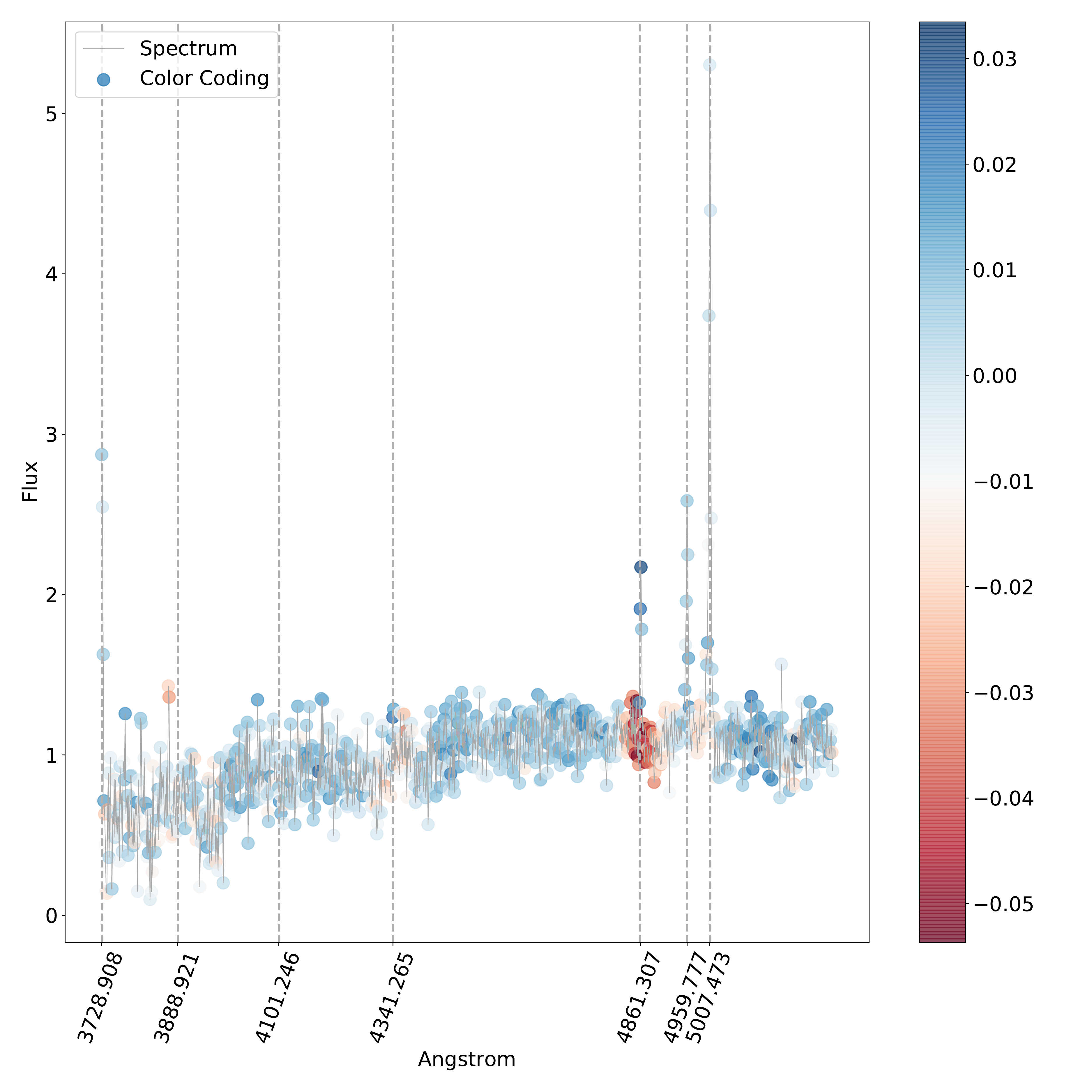}} \quad
\subfloat[Type 2, correctly classified]
{\label{fig:12corr}
\includegraphics[width=0.3\linewidth]{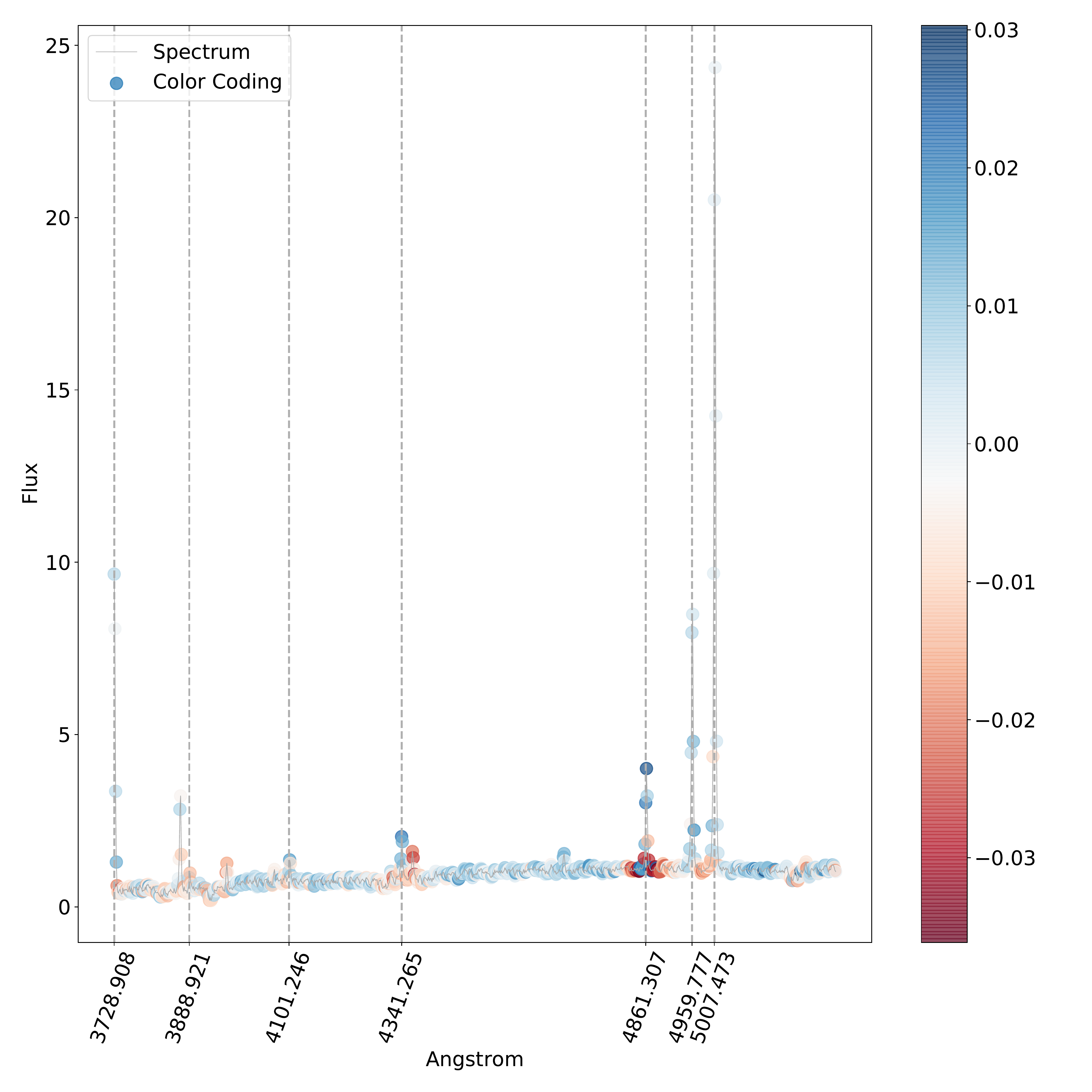}} \quad
\subfloat[Int. type, correctly classified]
{\label{fig:13corr}
\includegraphics[width=0.3\linewidth]{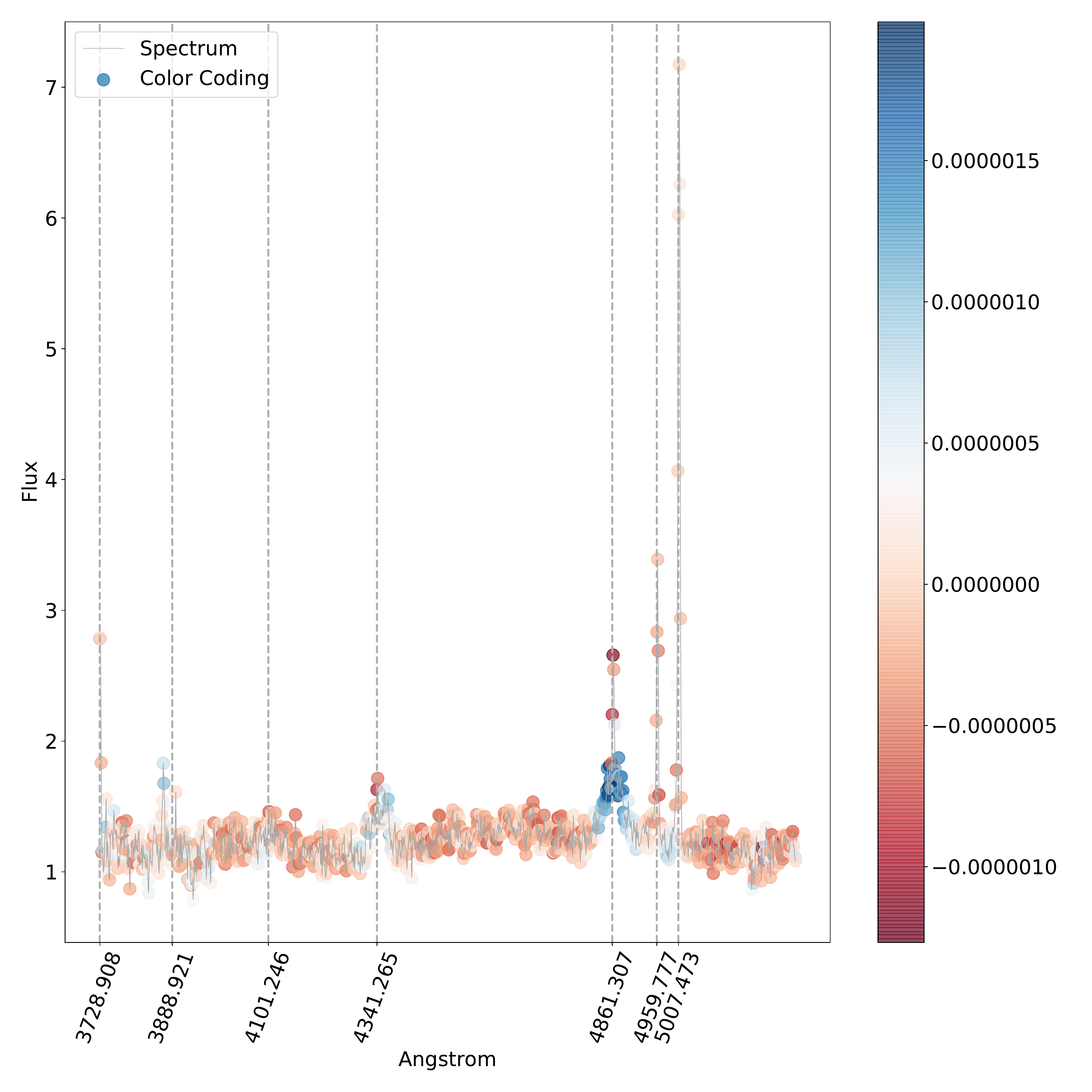}} \quad
\subfloat[Type 1, correctly classified]
{\label{fig:14corr}
\includegraphics[width=0.3\linewidth]{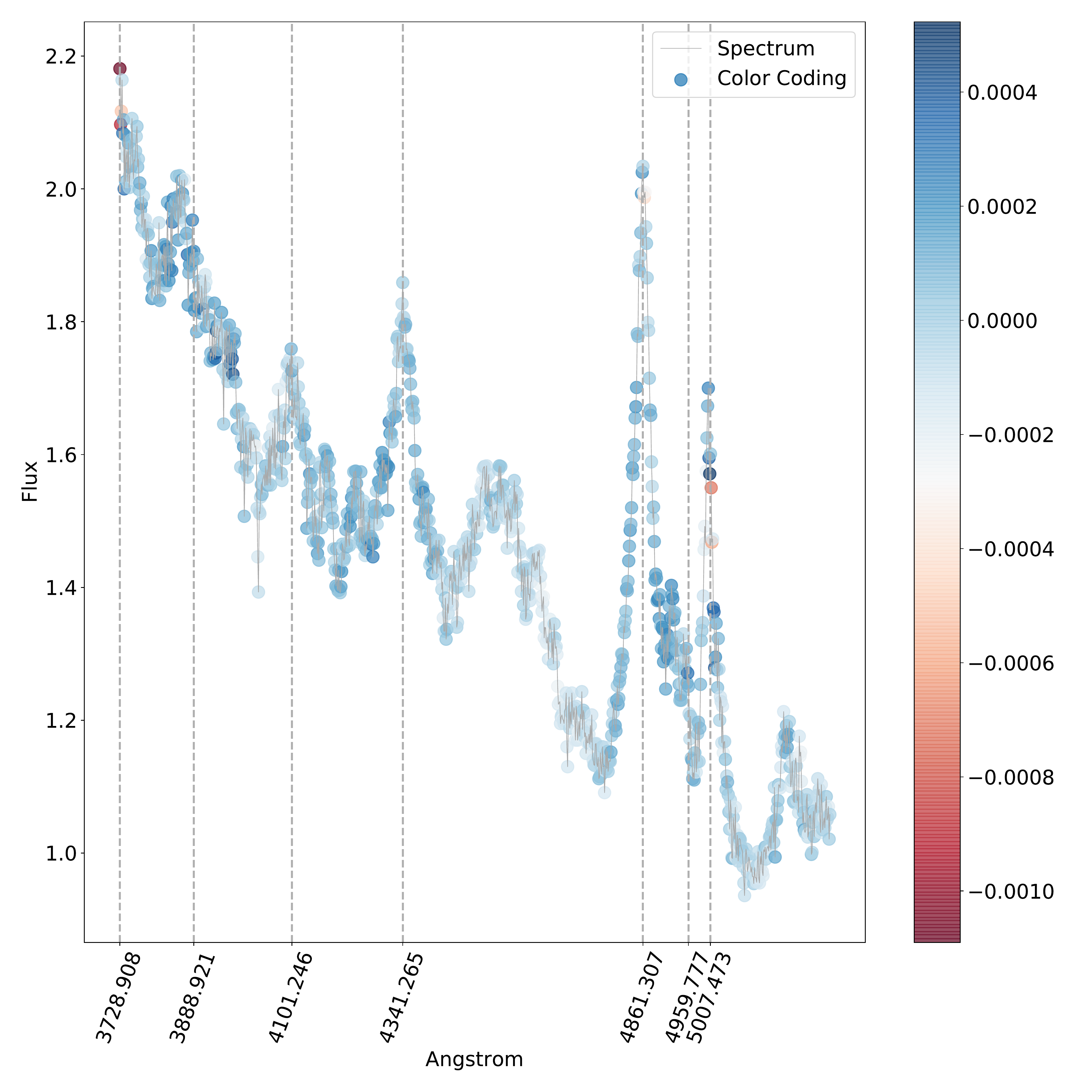}} \quad
\subfloat[Type 1, correctly classified]
{\label{fig:15corr}
\includegraphics[width=0.3\linewidth]{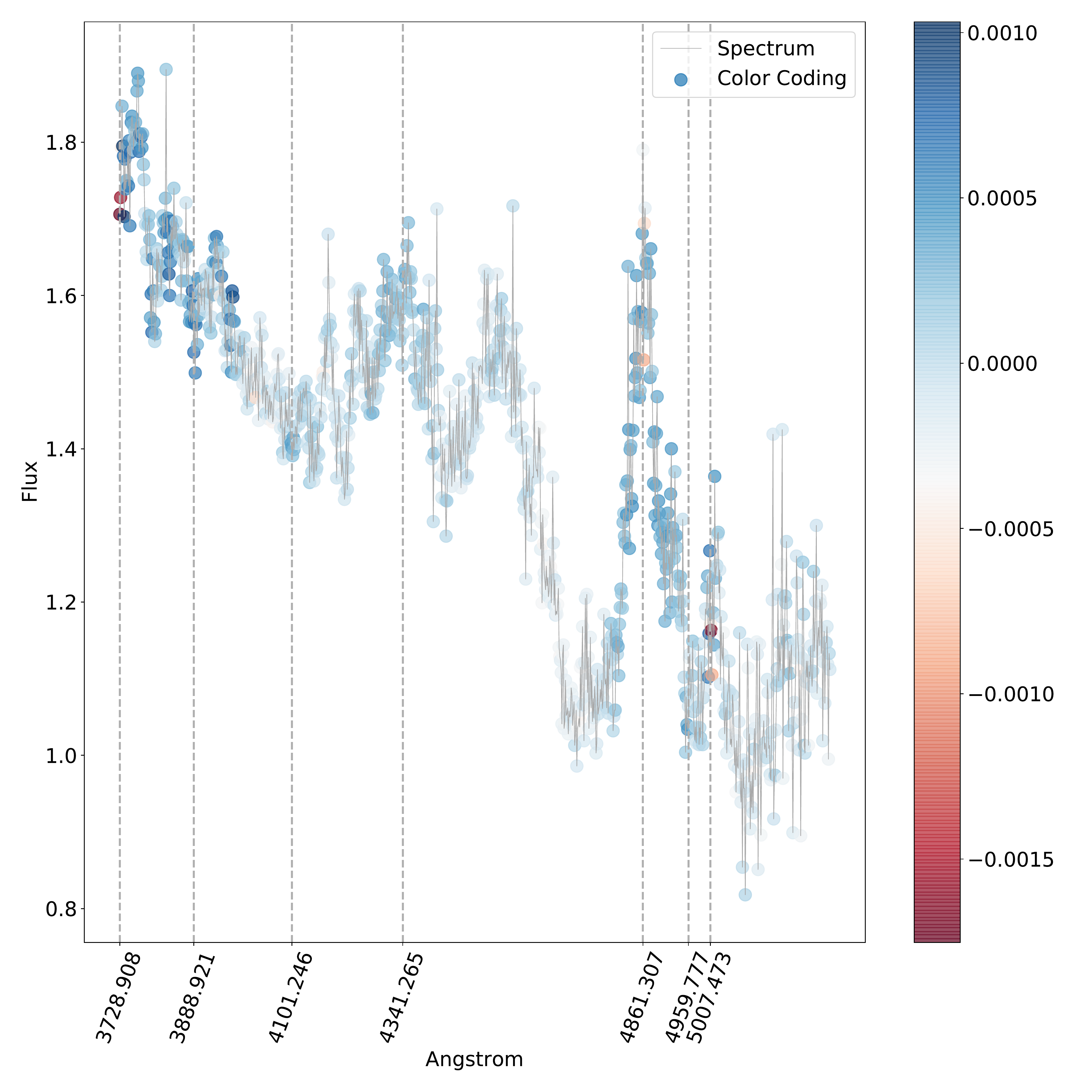}} \quad
\caption[Saliency maps for correctly classified spectra.]{Saliency maps for correctly classified spectra. \REFEREE{Color coding and axes as in Fig. ~\ref{fig:sal_maps_miss}.}}\label{fig:sal_maps_corr}
\end{figure*}

{In Fig.~\ref{fig:tsneannot} we show the spectra for which we calculated a saliency map projected onto the t-SNE Embedded Plane}. In the figure the red panels correspond to missclassified spectra and the green ones to correctly classified spectra. The corresponding saliency maps can be seen in Fig.~\ref{fig:sal_maps_miss} for missclassified spectra, and in Fig.~\ref{fig:sal_maps_corr} for correctly classified spectra. The numbering corresponds to the one reported in Fig.~\ref{fig:tsneannot}.

For spectra in Fig.~\ref{fig:sal_maps_miss} where an intermediate had been misclassified as Type 2, the cause of misclassification as inferred from the saliency map is the same as discussed above for Fig.~\ref{fig:second_salmap}. When the opposite misclassification occurs, we note that the $H \beta$ line appears often embedded in the underlying stellar absorption, a situation that is likely not common enough in our training set for the model to learn to deal properly with it.



Classification probabilities calculated by the SVM classifier can be seen for misclassified spectra in Table \ref{tab:miss_prob} and for correctly classified spectra in Table \ref{tab:corr_prob}.

The classification of spectra classified with an high confidence, like the 14-th spectrum in Fig.~\ref{fig:sal_maps_corr}, do not change considerably under small perturbations, as are the one used in this work to calculate the confidence derivative. This can be interpreted as the fact that single features, even if perturbed, do not change an high confidence prediction, {showing that} the results obtained with the SVM {are} robust.

\begin{table}
\caption[Classification probabilities for missclassified spectra for which saliency maps were calculated.]{Classification probabilities for missclassified spectra for which saliency maps were calculated. \REFEREE{Columns: reference index (first column), probabilities predicted for every class and ground truth in the last column. Rows: missclassified spectra highlighted in red in Fig. ~\ref{fig:tsneannot}}. \label{tab:miss_prob}}  
\centering
\begin{tabular}{lllll}
\hline \hline
Index & Prob. type 1 & Prob. int. & Prob.type 2 & True class \\ 
\hline
1 & 0.0 & 0.41 & 0.58 & Type 2 \\
2 & 0.0 & 0.68 & 0.32 & Type 2 \\
3 & 0.01 & 0.13 & 0.86 & Int. \\
4 & 0.0 & 0.48 & 0.52 & Type 2 \\
5 & 0.0 & 0.07 & 0.93 & Int. \\
6 & 0.0 & 0.14 & 0.86 & Int. \\
\hline
\end{tabular}
\end{table}

\begin{table}
\caption[Classification probabilities for missclassified spectra for which saliency maps were calculated.]{Classification probabilities for missclassified spectra for which saliency maps were calculated. \REFEREE{Columns and rows as in Tab. ~\ref{tab:miss_prob}}. \label{tab:corr_prob}}  
\centering
\begin{tabular}{lllll}
\hline \hline
Index & Prob. type 1 & Prob. int. & Prob.type 2 & True class \\ 
\hline
 7 & 0.99 & 0.01 & 0.00 & Type 1 \\
 8 & 0.00 & 0.02 & 0.98 & Type 2 \\
 9 & 0.00 & 0.95 & 0.05 & Int. \\
10 & 0.00 & 0.05 & 0.95 & Type 2 \\
11 & 0.00 & 0.05 & 0.95 & Type 2 \\
12 & 0.00 & 1.00 & 0.00 & Int. \\
13 & 1.00 & 0.00 & 0.00 & Type 1 \\
14 & 0.99 & 0.01 & 0.00 & Type 1 \\
\hline
\end{tabular}
\end{table}

\section{Conclusions}
We trained a support-vector machine model to classify AGN spectra, obtaining fairly accurate results on a test set not seen in training (F-score of $\approx94\%$). While it is tempting to just apply the trained model to a large sample of spectra, we argue that it is crucial to first understand why the classifier returns the prediction it does. We have shown that simple interpretability tools, such as a saliency map, allow us to easily accomplish this, at least on a spectrum-by-spectrum basis. Even though a general explanation of the criteria used by a classifier (as would be achieved by some natively interpretable machine learning method) is in general impossible to achieve for a black box classifier, saliency maps make it possible to understand the workings of an otherwise black-box classifier in the neighborhood of any given datapoint.

We computed saliency maps of a random sample of correctly classified and misclassified spectra. In general we find that the regions of the spectrum that most affect the classifier prediction are similar to those used by a human expert, i.e. those around the spectral lines $[O II] 3727$, $He I 3889$, $H \delta 4101$, $H \gamma 4340$, $H \beta 4861$, $[O III] 4959$ and $[O III] 5007$. Also the way in which the model uses the information in these regions conforms to our expectations: for example it implicitly relies on the width of the $H \beta$ line which increases the probability of classifying a spectra as type 1.
We thus conclude that, at least on the spectra we considered, our classifier operates pretty much in the same way as a human would, just automatically and much faster. This is extremely reassuring regarding the possibility of applying machine learning classifiers to the large datasets of spectra that will result from  upcoming surveys, which will not be amenable to direct human classification.

We also visualized the high-dimensional feature space of the spectra using the t-SNE algorithm, which maps spectra to points in a plane while attempting to preserve the local pairwise distances. We find that type 1 and type 2 spectra are mapped to distinct regions of the plane, forming two 'islands' separated by a clear-cut isthmus. If intermediate-type spectra are also included, some of them happen to populate the isthmus, forming a bridge between type 1 and type 2, as expected from the very definition of intermediate type. However, several intermediate-type spectra end up in the same region occupied by type 2 spectra, apparently mixed with them. It may be that labeling these spectra as intermediate-type is questionable in the first place. Interestingly, both intermediate-type and type-2 spectra show subclustering structure in the t-SNE plane. While this may be an artifact of t-SNE, it persists when different values of the perplexity hyperparameter are used (perplexity roughly corresponds to the expected size of groups in the dataset) which suggests that the result is genuine. Further work is needed to characterize these subgroups, perhaps comparing them with proposed AGN subtypes; we plan to carry this out in a subsequent paper.

\begin{acknowledgements}
This project has received funding from the European Union's Horizon $2020$
research and innovation program under the Marie Sk\l{}odowska-Curie grant agreement No. $664931$. {This material is based upon work supported by Tamkeen under the NYU Abu Dhabi Research Institute grant CAP3}.\\

Funding for the Sloan Digital Sky Survey IV has been provided by the Alfred P. Sloan Foundation, the U.S. Department of Energy Office of Science, and the Participating Institutions. SDSS-IV acknowledges
support and resources from the Center for High-Performance Computing at
the University of Utah. The SDSS web site is www.sdss.org.

SDSS-IV is managed by the Astrophysical Research Consortium for the 
Participating Institutions of the SDSS Collaboration including the 
Brazilian Participation Group, the Carnegie Institution for Science, 
Carnegie Mellon University, the Chilean Participation Group, the French Participation Group, Harvard-Smithsonian Center for Astrophysics, 
Instituto de Astrof\'isica de Canarias, The Johns Hopkins University, Kavli Institute for the Physics and Mathematics of the Universe (IPMU) / 
University of Tokyo, the Korean Participation Group, Lawrence Berkeley National Laboratory, 
Leibniz Institut f\"ur Astrophysik Potsdam (AIP),  
Max-Planck-Institut f\"ur Astronomie (MPIA Heidelberg), 
Max-Planck-Institut f\"ur Astrophysik (MPA Garching), 
Max-Planck-Institut f\"ur Extraterrestrische Physik (MPE), 
National Astronomical Observatories of China, New Mexico State University, 
New York University, University of Notre Dame, 
Observat\'ario Nacional / MCTI, The Ohio State University, 
Pennsylvania State University, Shanghai Astronomical Observatory, 
United Kingdom Participation Group,
Universidad Nacional Aut\'onoma de M\'exico, University of Arizona, 
University of Colorado Boulder, University of Oxford, University of Portsmouth, 
University of Utah, University of Virginia, University of Washington, University of Wisconsin,
Vanderbilt University, and Yale University.

The code on which this work is based can be found at the following link \url{https://gitlab.com/tobia.peruzzi/agn_spectra}
\end{acknowledgements}

%
%

\bibliographystyle{aa}
\bibliography{manuscript}

\appendix
\section{Effects of feature scaling and perplexity on t-SNE results}

\begin{figure}
\includegraphics[scale=0.2]{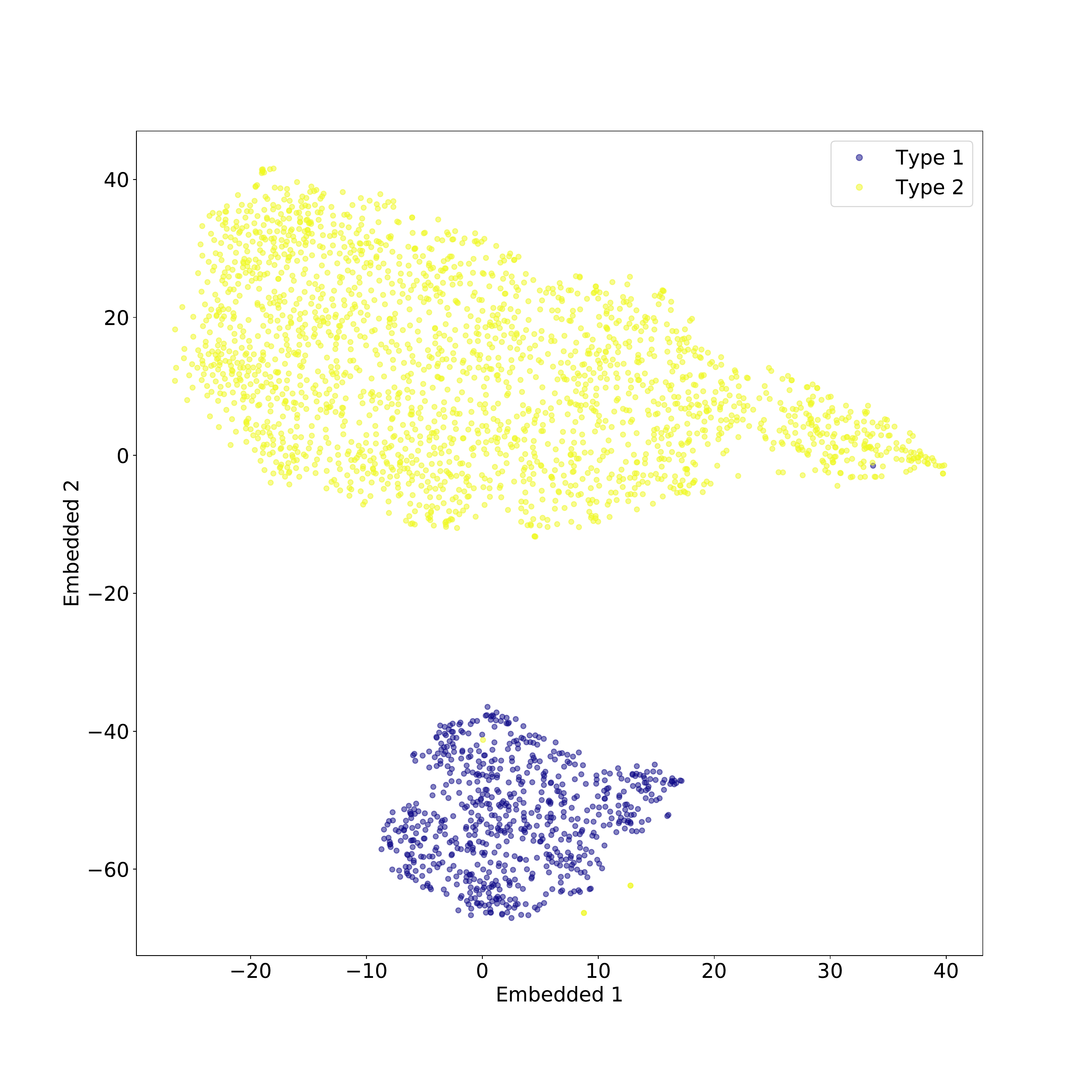}
\caption{t-SNE embedded plane for type 1 and type 2 AGN. \REFEREE{Blue points: type 1. Yellow points: type 2.}}
\label{fig:tsne_1_2}
\end{figure}

\begin{figure}[bth]
\subfloat[Perplexity: 5]
{\label{fig:t_sne_p5}
\includegraphics[width=.45\linewidth]{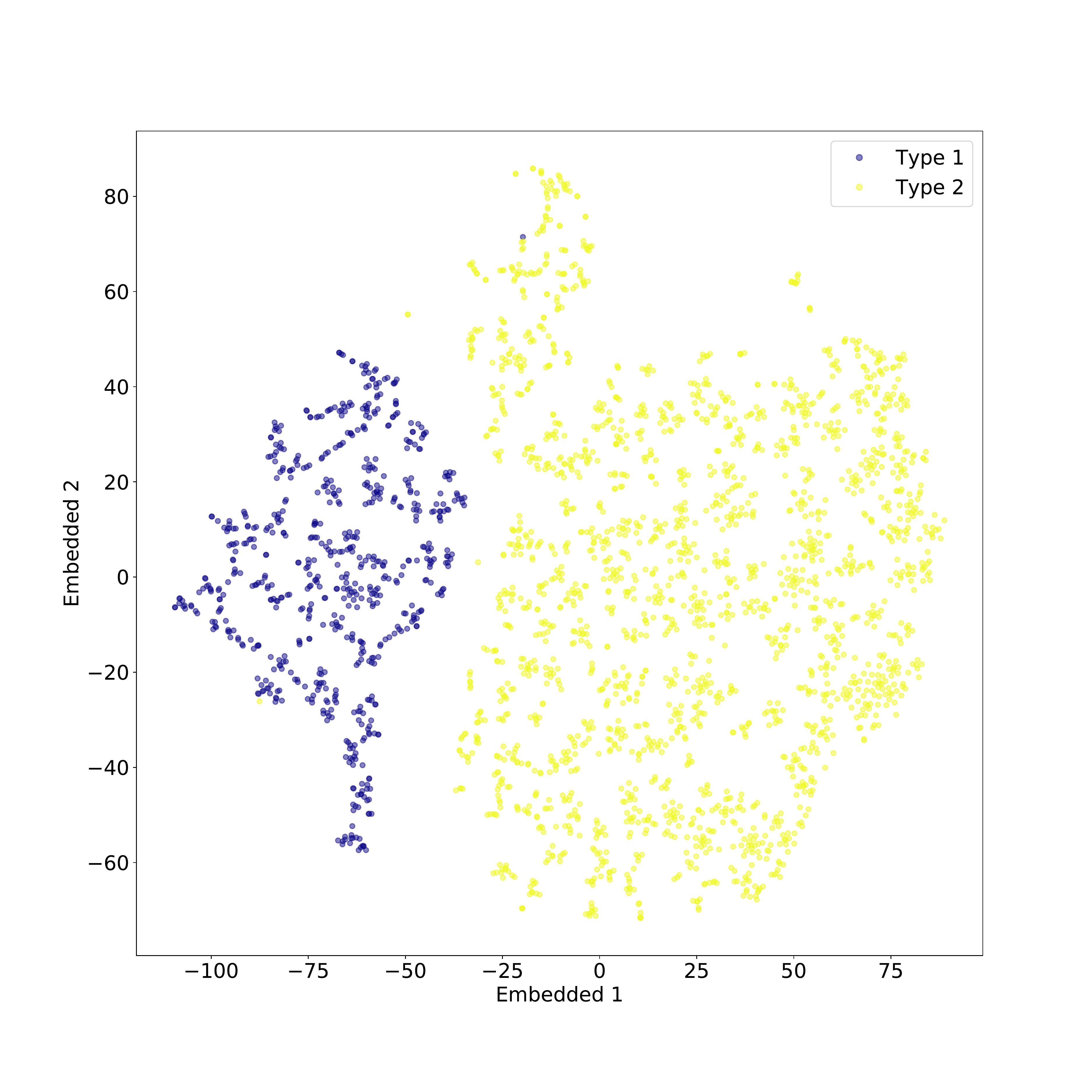}} \quad
\subfloat[Perplexity: 15]
{\label{fig:t_sne_p15}
\includegraphics[width=.45\linewidth]{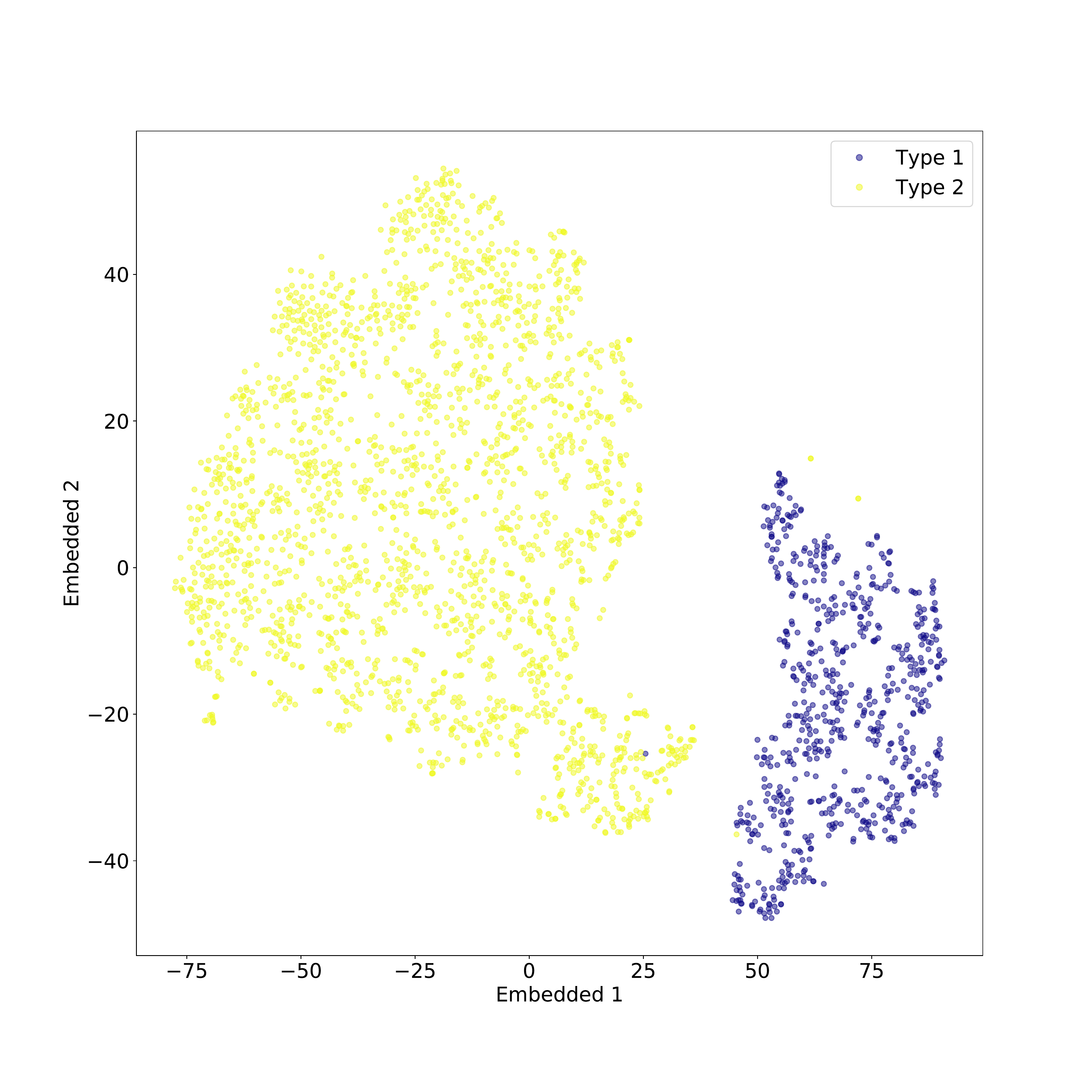}} \\
\subfloat[Perplexity: 30]
{\label{fig:t_sne_p30}
\includegraphics[width=.45\linewidth]{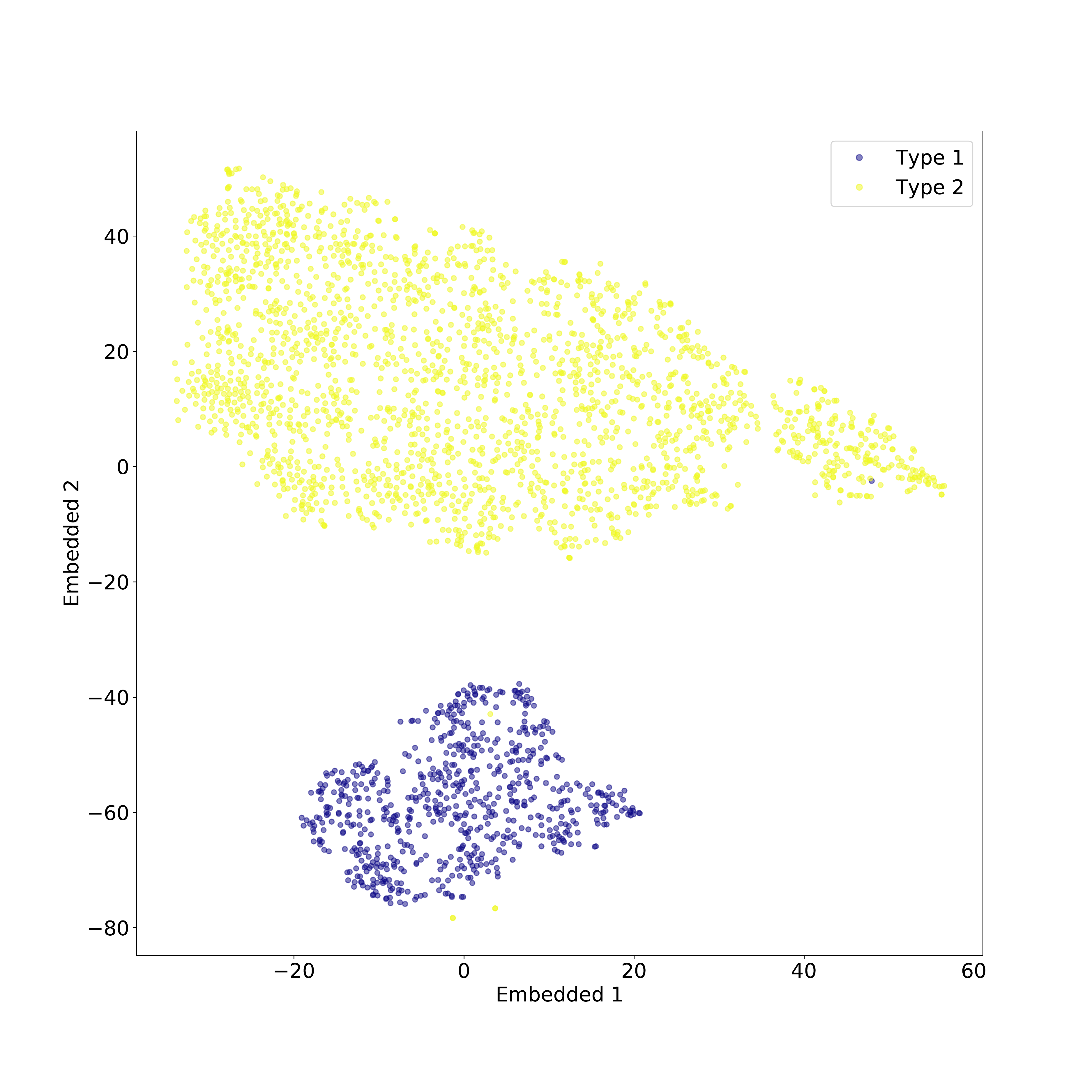}} \quad
\subfloat[Perplexity: 40]
{\label{fig:t_sne_p40}
\includegraphics[width=.45\linewidth]{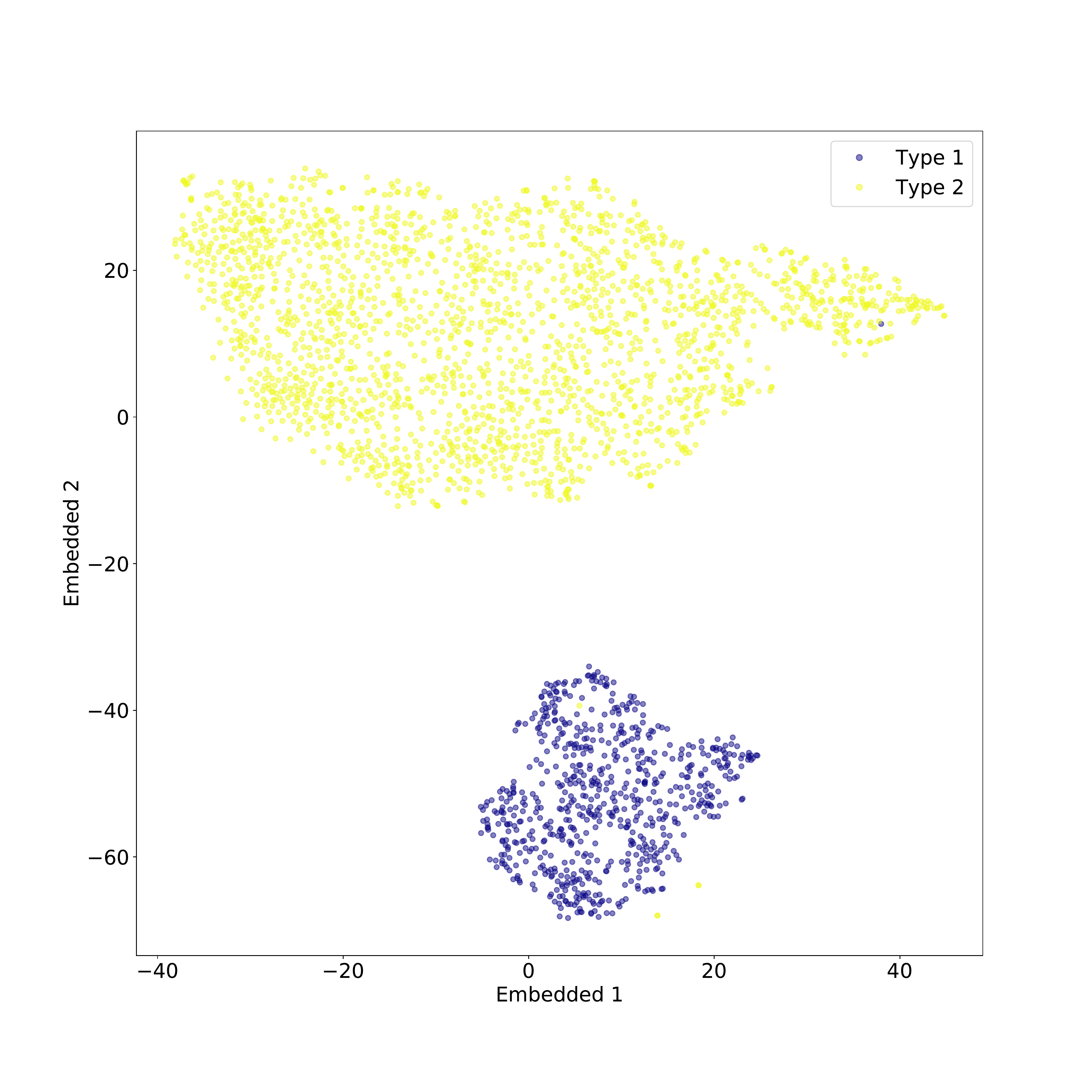}}
\caption[t-SNE for type 1 and type 2 with various perplexity values.]{t-SNE for type 1 and type 2 with various perplexity values. \REFEREE{Color points as in Fig. ~\ref{fig:t_sne_perplexities}.}\label{fig:t_sne_perplexities}}
\end{figure}

\begin{figure}[bth]
\subfloat[Perplexity: 5]
{\label{fig:t_sne_p5_complete}
\includegraphics[width=.45\linewidth]{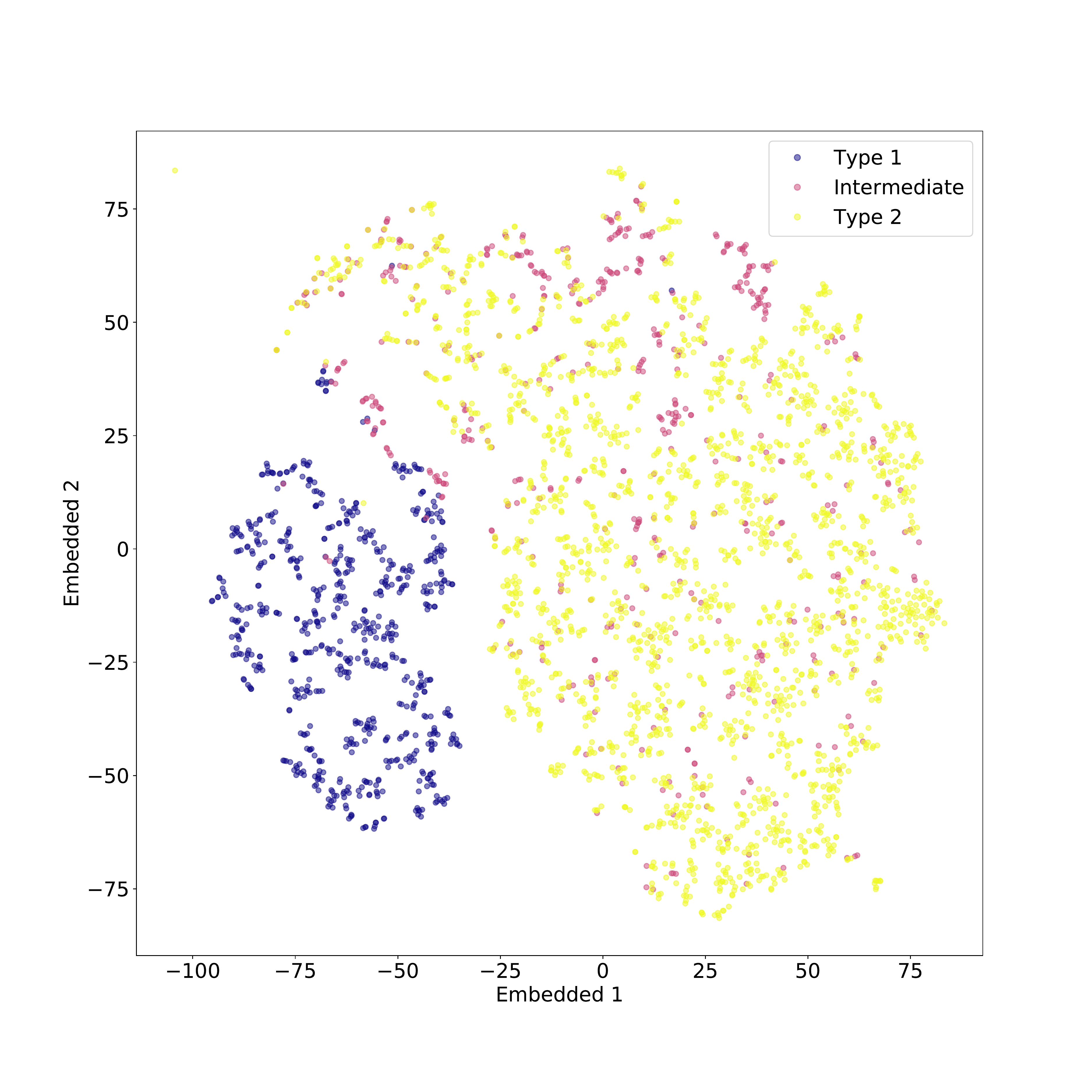}} \quad
\subfloat[Perplexity: 15]
{\label{fig:t_sne_p15_complete}
\includegraphics[width=.45\linewidth]{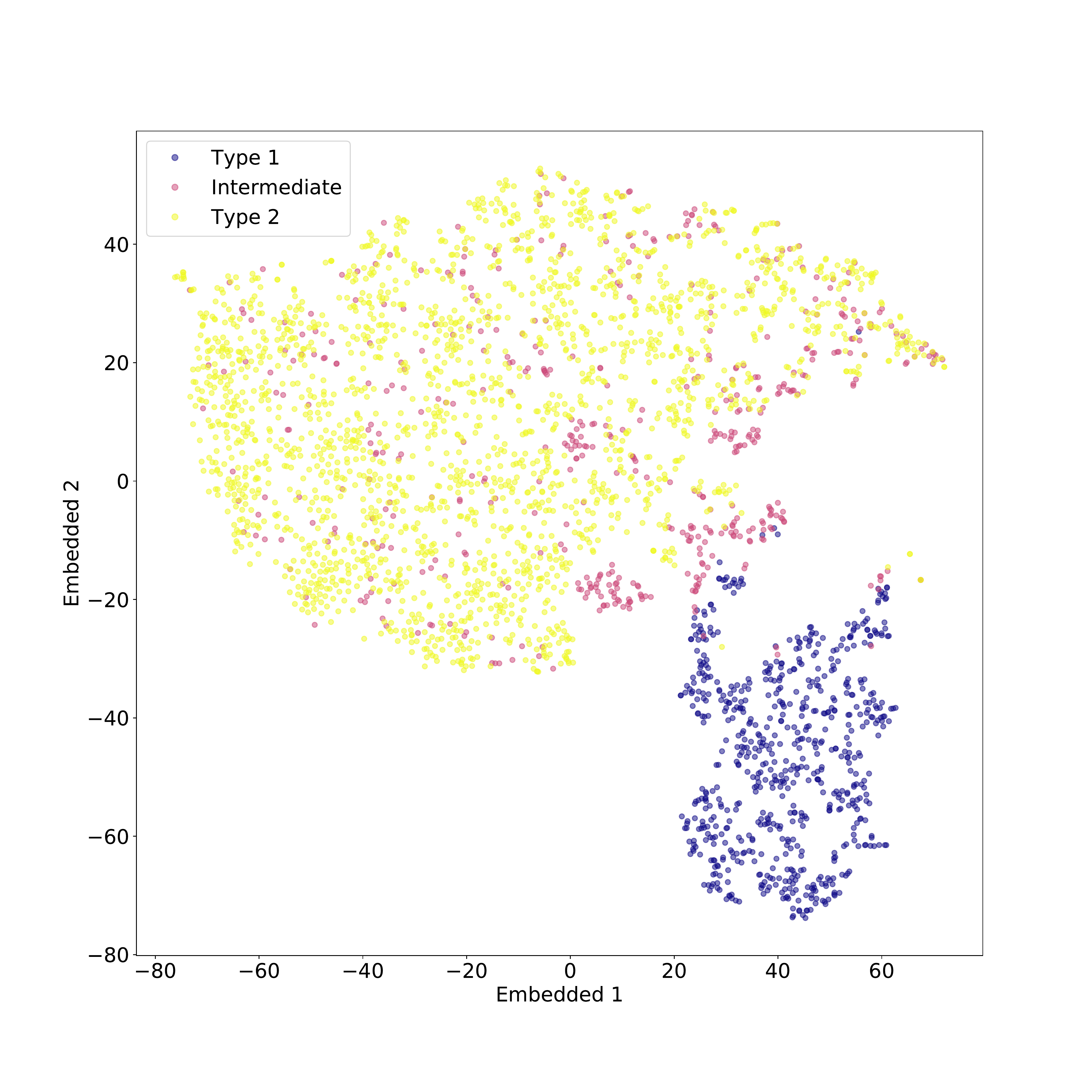}} \\
\subfloat[Perplexity: 30]
{\label{fig:t_sne_p30_complete}
\includegraphics[width=.45\linewidth]{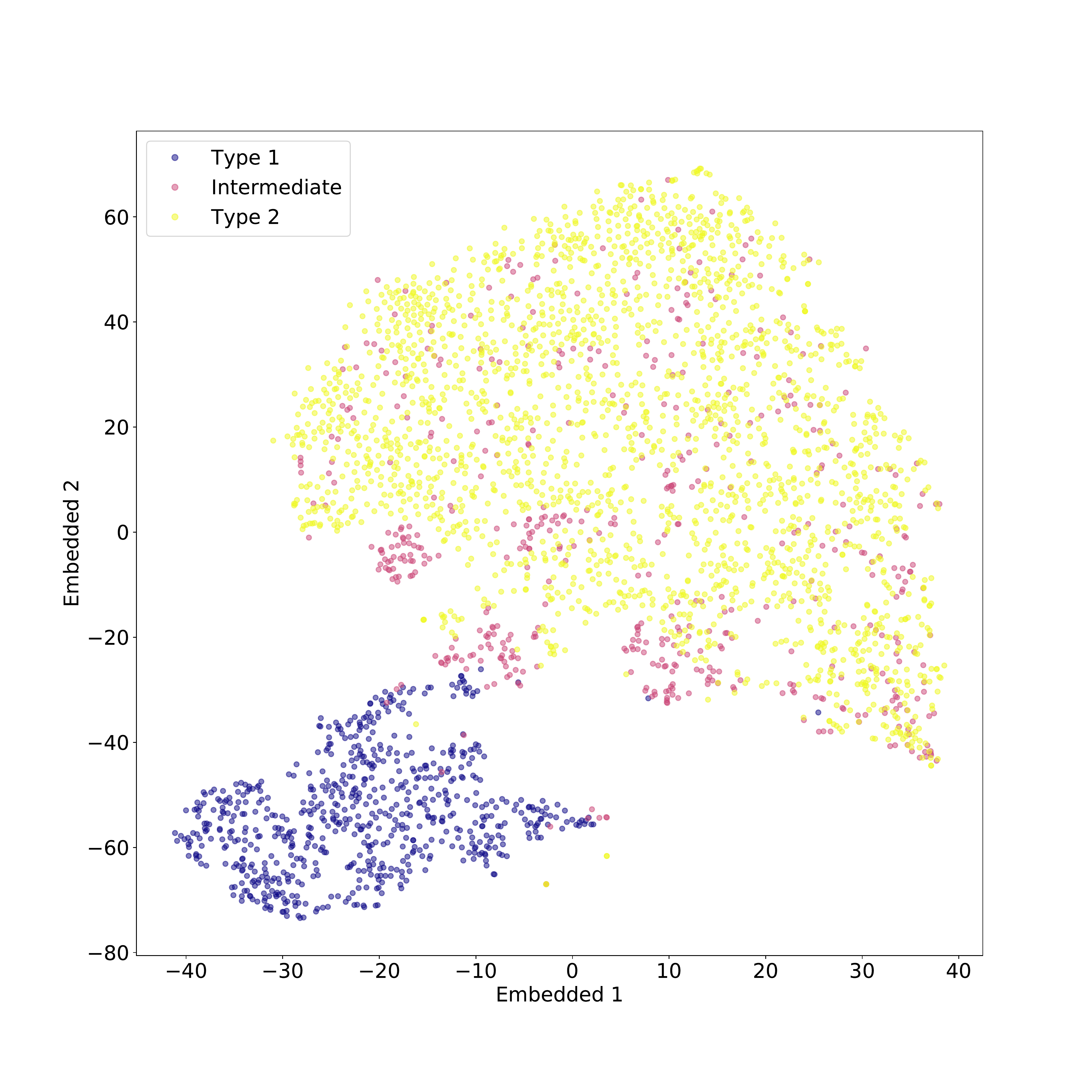}} \quad
\subfloat[Perplexity: 40]
{\label{fig:t_sne_p40_complete}
\includegraphics[width=.45\linewidth]{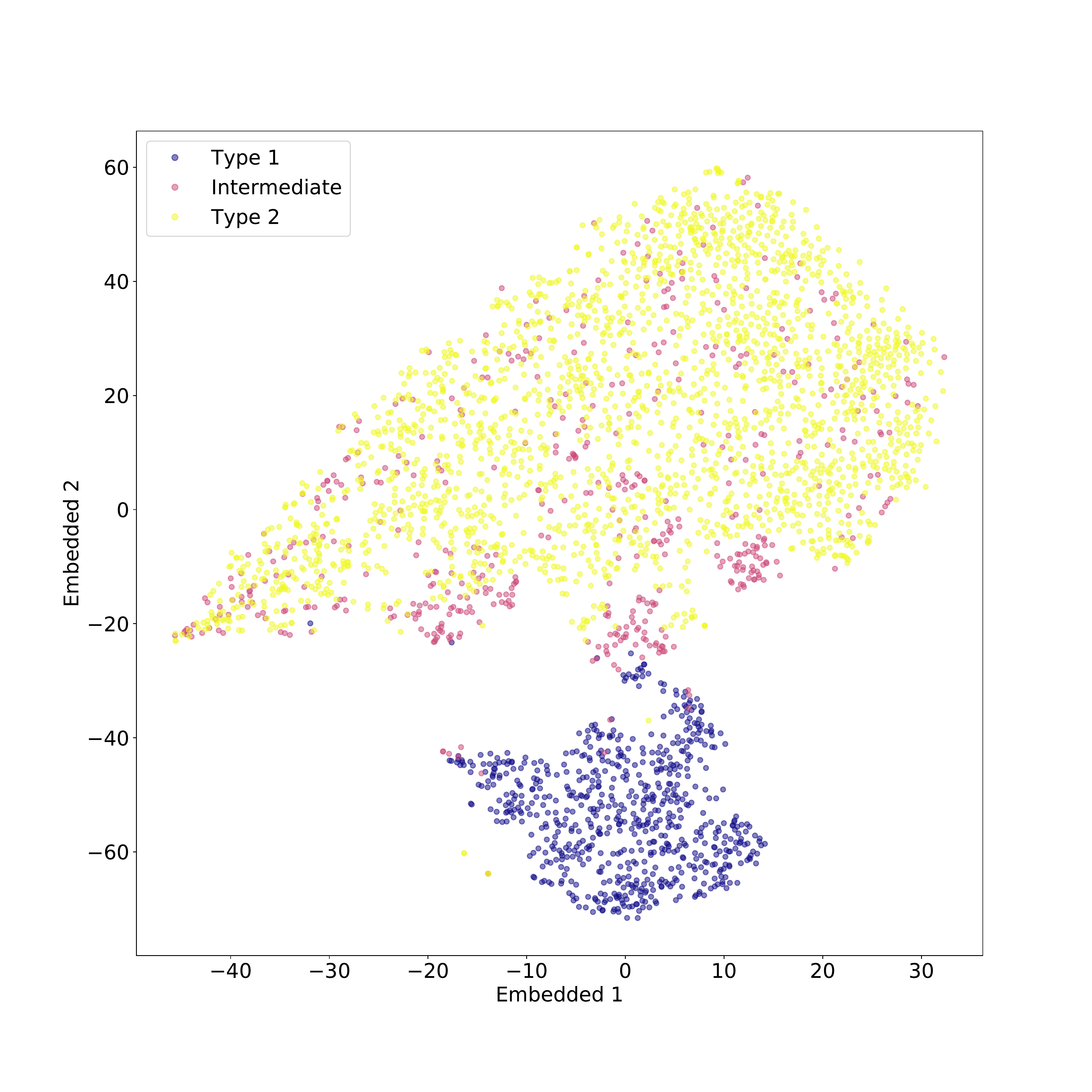}}
\caption[t-SNE for whole dataset with various perplexity values.]{t-SNE for whole dataset with various perplexity values. \REFEREE{Color points as in Fig. \ref{fig:tsne_tot}.}} \label{fig:t_sne_perplexities_whole}
\end{figure}

The t-SNE algorithm depends on a tunable parameter, perplexity, which loosely corresponds to the expected number of neighbors of the typical point in the dataset under consideration. The visualization produced by t-SNE can vary strongly as perplexity is changed and there is no general rule on how to pick the right value of this parameter. This may result in misleading visualizations, so it is best to try different values of perplexity and be wary of features (e.g. data subclusters) that only show up in a narrow range of perplexities \citep[][]{wattenberg2016how}. In Fig.~\ref{fig:t_sne_perplexities} we explore the effects of varying perplexity between $5$ and $40$ for type 1 and type 2 AGNs, while in Fig.~\ref{fig:t_sne_perplexities_whole} we also include intermediate-type AGNs.

\begin{figure}
\includegraphics[scale=0.2]{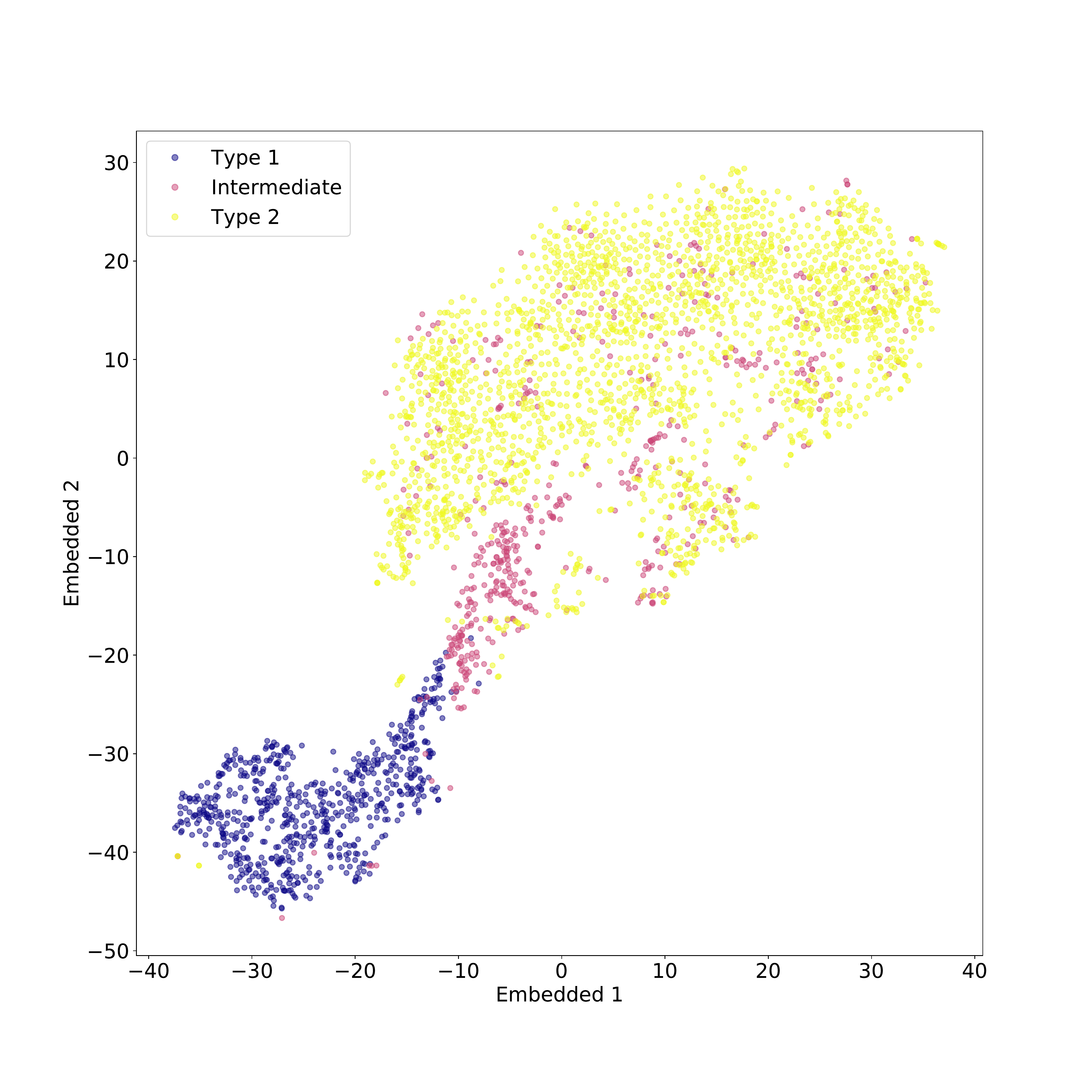}
\caption{t-SNE embedded plane for type 1, type 2 and intermediate type AGN spectra scaled and mean normalized. \REFEREE{Color points as in Fig. \ref{fig:tsne_tot}.}}
\label{fig:tsne_tot_scaled}
\end{figure}

\begin{figure}[bth]
\subfloat[Perplexity: 5]
{\label{fig:t_sne_p5_complete_scal}
\includegraphics[width=.45\linewidth]{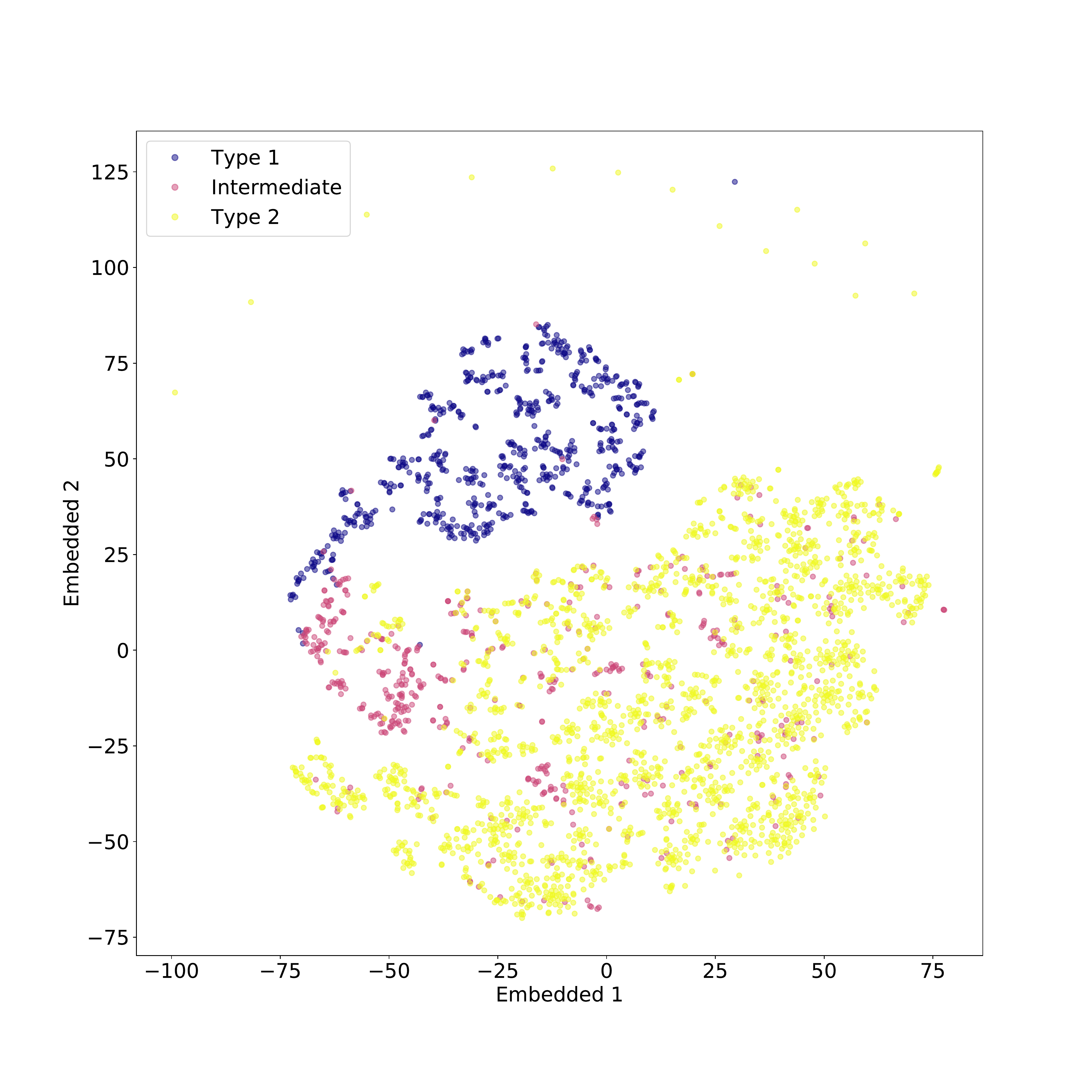}} \quad
\subfloat[Perplexity: 15]
{\label{fig:t_sne_p15_complete_scal}
\includegraphics[width=.45\linewidth]{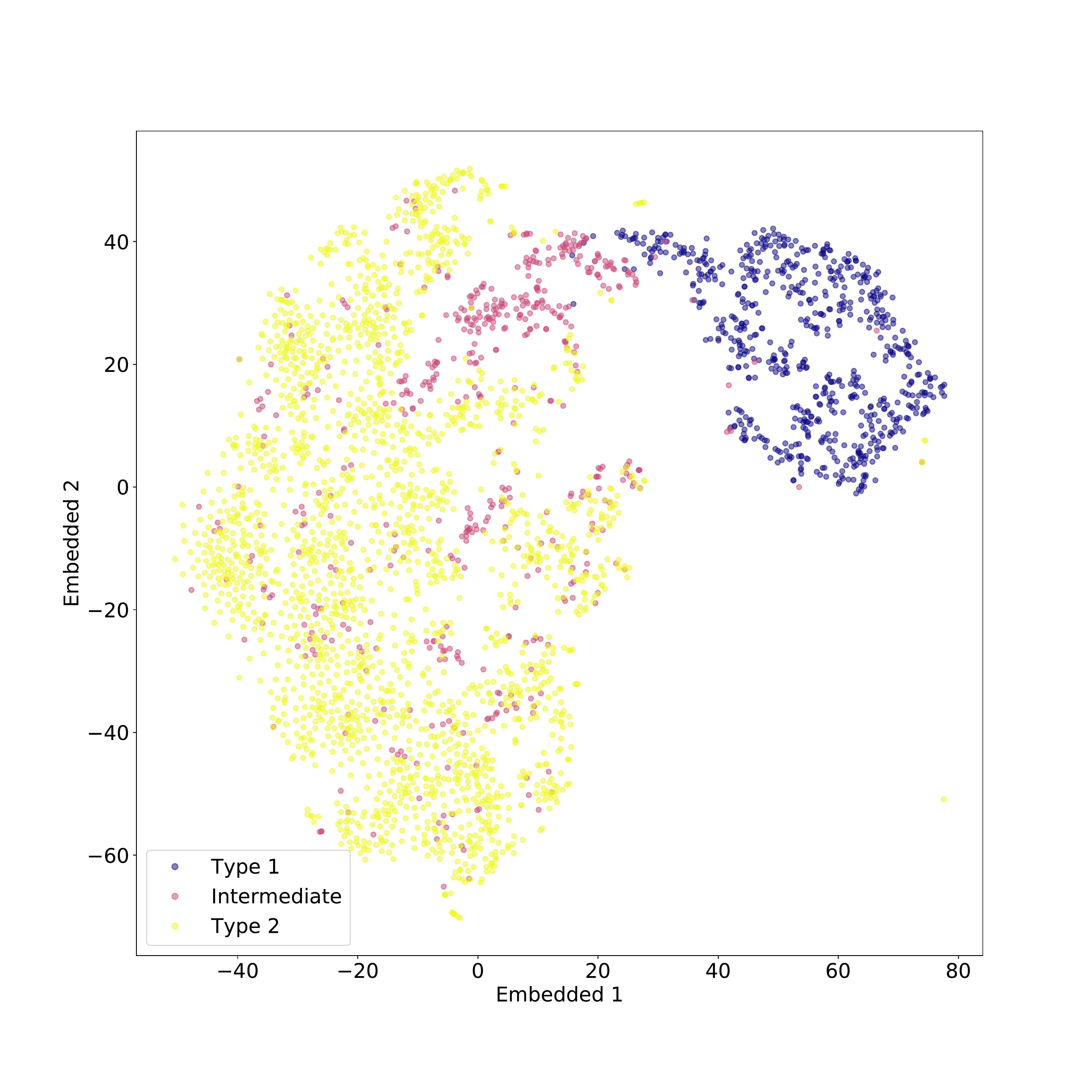}} \\
\subfloat[Perplexity: 30]
{\label{fig:t_sne_p30_complete_scal}
\includegraphics[width=.45\linewidth]{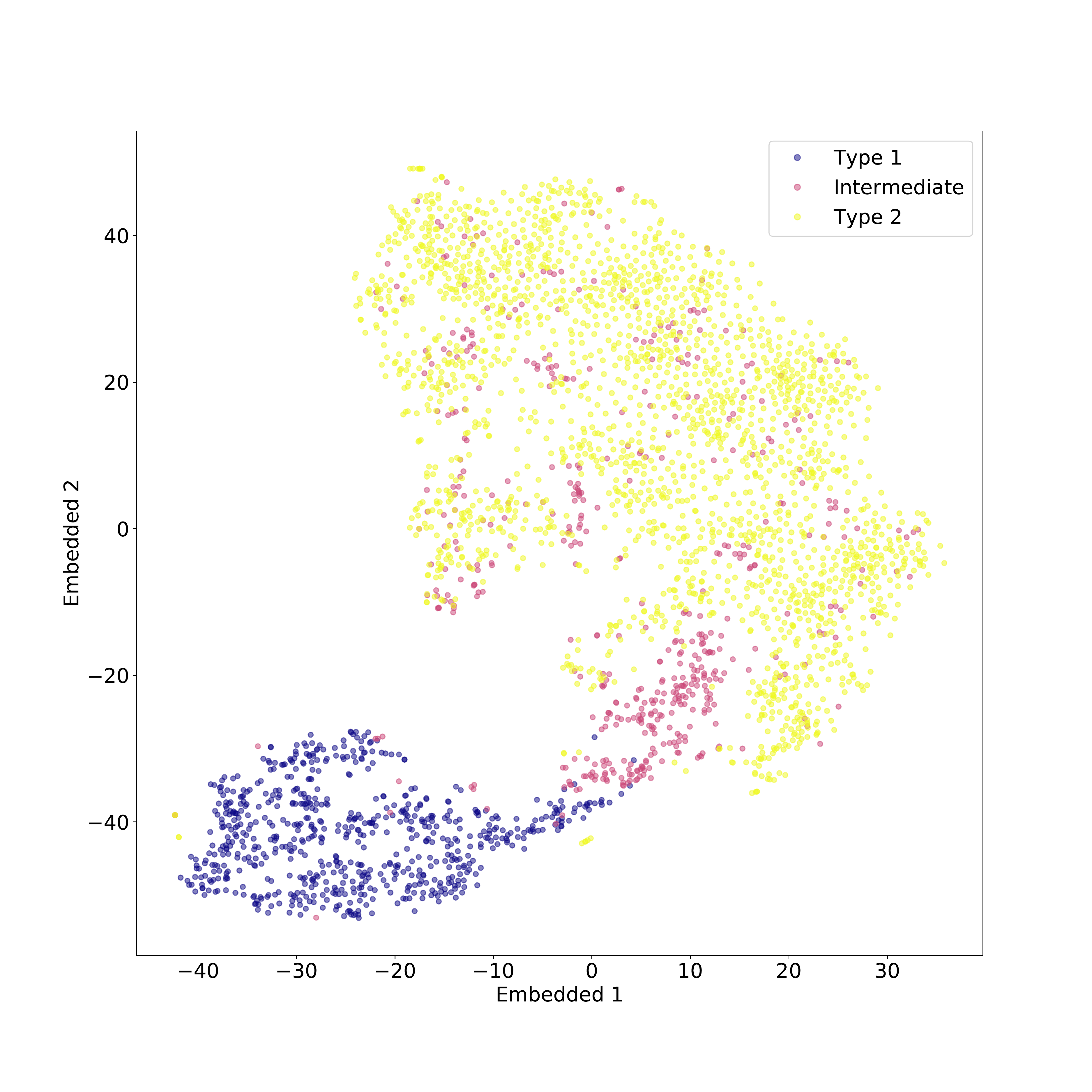}} \quad
\subfloat[Perplexity: 40]
{\label{fig:t_sne_p40_complete_scal}
\includegraphics[width=.45\linewidth]{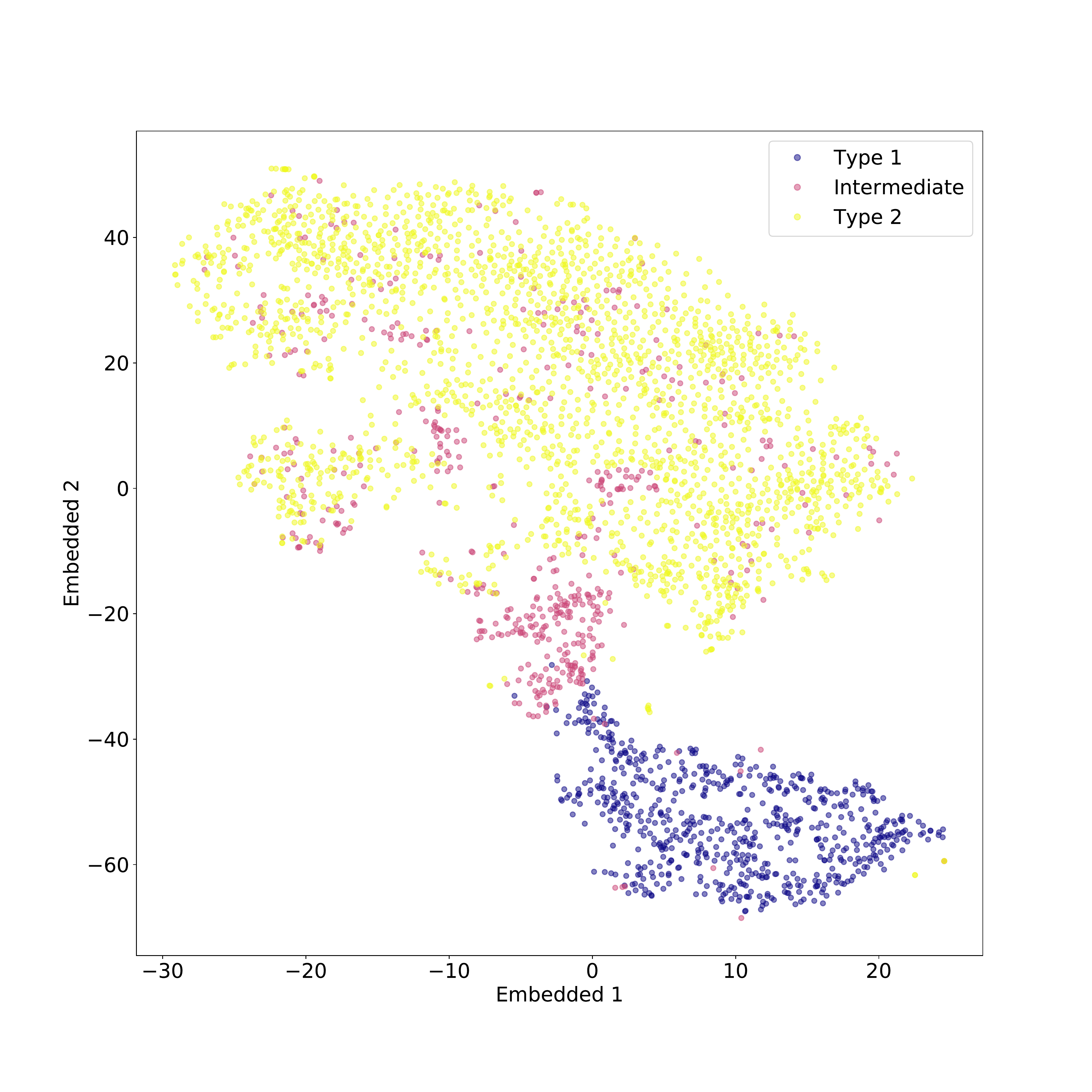}}
\caption[t-SNE for whole dataset scaled and mean normalized with various perplexity values.]{t-SNE for whole dataset scaled and mean normalized with various perplexity values. \REFEREE{Color points as in Fig. \ref{fig:tsne_tot}.}}\label{fig:t_sne_perplexities_whole_scal}
\end{figure}

After that, the t-SNE algorithm was fed with scaled and mean normalized data, that is every feature $x_{i}$ is expressed by:
\begin{equation}
    x_{i} = \frac{x_{i} - \mu_{i}}{s_{i}}
\end{equation}
where $\mu_{i}$ is the average of the i-th feature and $s_{i}$ is the standard deviation of the i-th feature. The results can be seen in Fig.~ \ref{fig:tsne_tot_scaled}, with a perplexity of 50. The result for other values of the perplexity can be seen in Fig.~\ref{fig:t_sne_perplexities_whole_scal}.

Overall the outcome is similar to the un-normalized case, and t-SNE (with the right choice of perplexity) seems to perform just as well after feature scaling and mean normalization. 
The result presented in Figure \ref{fig:tsne_tot_scaled} can be interpreted even more clearly as a transition from type 1 spectra, characterized by broad lines and strong continuum, to intermediate spectra, characterized by narrower lines and lower continuum, and from intermediate to type 2 spectra, characterized by narrow lines and almost constant continuum. Of course some intermediate spectra will still be clustered together with type 1, or more often with type 2 spectra, but this is an expected result. In fact the distinction between intermediate and type 2 spectra is not strict and spectra of the two types may appear similar. Nonetheless the figure shows a clear transition region between type 1 and type 2 populated by intermediate-type spectra.


\end{document}